\documentstyle[prb,preprint,eqsecnum,aps]{revtex}
\tighten
\begin{document}
\draft
\title{Time-Dependent Ginzburg-Landau Equations for Mixed $d$- and $s$-Wave  
Superconductors}
\author{Jian-Xin Zhu, Wonkee Kim, and 
C. S. Ting} 
\address{Texas Center for Superconductivity, University of Houston, 
Houston, Texas 77204}
\author{Chia-Ren Hu}
\address{Department of Physics, Texas A\&M University, College Station, 
Texas 77843}

%\date{\today}
\maketitle
\begin{abstract}
A set of coupled time-dependent Ginzburg-Landau equations (TDGL) for 
superconductors of mixed $d$-  and $s$-wave symmetry are derived 
microscopically from the Gor'kov equations by using the analytical 
continuation technique. The scattering effects due to impurities 
with both nonmagnetic and magnetic interactions are considered.
We find that the $d$- and $s$-wave components of the order parameter 
can have very different relaxation times in the presence of nonmagnetic 
impurities. This result is contrary to a set of phenomenologically 
proposed TDGL equations and thus may lead to new physics in the 
dynamics of flux motion. 
\end{abstract}

\narrowtext

\section{Introduction}
There are growing experimental evidences to suggest that high-$T_{c}$ 
superconductors have a dominant $d_{x^{2}-y^{2}}$-wave pairing 
symmetry.~\cite{Harlingen95} Based on symmetry considerations, 
Volovik~\cite{Volovik93} argued that an $s$-wave component of 
the order parameter should be generated near the core region of a 
vortex in a $d$-wave superconductor.  This conclusion was later 
confirmed by a numerical calculation~\cite{SKB94} and by studying 
a set of microscopically-derived two-component 
Ginzburg-Landau (GL) equations.~\cite{RXT95}  

In view of the enormous success of the GL theory 
for describing the equilibrium properties of superconductors near 
$T_{c}$, it is natural to generalize it to time-dependent 
situations. This generalization has become particularly desirable, 
since a set of phenomenological time-dependent Ginzburg-Landau (TDGL) 
equations for coupled $s$- and $d$-wave superconducting order parameters 
has been recently proposed,~\cite{ADB97} and used to 
investigate the dynamics of vortices in high-$T_c$ superconductors. 
One would very much like to know how valid is such an approach.

It is well known, however, that TDGL equations are not as universal 
in form as its time-independent variety, but can be dependent strongly 
on whether the system is gapful or gapless, and in the later case, 
whether a strong or weak gaplessness condition is assumed.
The simplest set of TDGL equations for conventional $s$-wave 
superconductors was proposed phenomenologically by 
Schmid,~\cite{Schmid66} and subsequently derived  
microscopically by Gor'kov and \'{E}liashberg~\cite{Gor68} 
under the assumption of a strong gaplessness condition ({\it i.e.},
$\tau_sT_c \ll 1$, where $\tau_s$ is the spin-flip lifetime and 
$T_c$ is the transition temperature). 
This set of equations has been used in the past to study the 
vortex dynamics in conventional superconductors.~\cite{Hu72} 
\'{E}liashberg~\cite{Eliash69} has later derived a more complex set of 
TDGL equations for low-$T_c$ $s$-wave superconductors assuming only 
the weak gaplessness condition [{\it i.e.}, $\tau_s\Delta_0 \ll 1$, 
where $\Delta_0(T)$ is the equilibrium value of the ($s$-wave) order 
parameter in the absence of fields]. It has been used to study flux-flow 
resistivity~\cite{Hu73} and the transport entropy of vortices.~\cite{Hu76}
Even more complex sets of TDGL equations have been derived subsequently
assuming only the dirty limit condition ($\tau_1 T_c \ll 1$, where $\tau_1$
is the total scattering lifetime) and 
$(1-T/T_c)\ll 1$,~\cite{Hu80} so the system need no longer be gapless.
However,  this set of equations is so complex that it has not yet been
extensively used.

In this work, we shall derive microscopically a set of coupled 
TDGL equations for superconductors with mixed $d$- and $s$-wave 
pairing symmetry based on the approach of Gor'kov and 
\'{E}liashberg~\cite{Gor68,Eliash69} in the presence of impurities with both 
spin-flip and non-spin-flip interactions and assuming only weak 
gaplessness conditions for both waves
({\it i.e.}, $\tau_1\Delta_{d0}\ll 1$, and $\tau_s\Delta_{s0}\ll 1$.)  
The primary objective of this derivation is to establish a reasonably 
reliable set of equations governing the dynamics of coupled
$d$- and $s$-wave order parameters which are hopefully valid for
describing the dynamic properties of high-$T_c$ superconductors.

The outline of this paper is as follows: In Sec. II, the TDGL equations 
for the order parameters are
derived. The expressions for current and charge 
density are presented in Sec. III. Finally, discussions and summary are 
given in Sec. IV. 

\section{Time-Dependent Ginzburg-Landau Equations for the Order Parameters} 
We begin with the Gor'kov 
equations:~\cite{Gor60}  
\begin{mathletters}
\label{EQ:motion}
\begin{eqnarray}
&[-\frac{\partial}{\partial \tau}-h({\bf x}\tau)]
G_{\alpha\beta}({\bf x}\tau, {\bf x}^{\prime}\tau^{\prime})
-U_{\alpha\gamma}({\bf x})
G_{\gamma\beta}({\bf x}\tau,{\bf x}^{\prime}\tau^{\prime})&\nonumber \\
&+ \int d{\bf x}^{\prime\prime}\Delta_{\alpha\gamma}
({\bf x}\tau^{0+},{\bf x}^{\prime\prime}\tau)F_{\gamma\beta}^{\dagger}
({\bf x}^{\prime\prime}\tau,{\bf x}^{\prime}\tau^{\prime})
=\delta({\bf x}-{\bf x}^{\prime})\delta(\tau-\tau^{\prime})
\delta_{\alpha\beta}\;,&\\
&[\frac{\partial }{\partial \tau}-h^{*}({\bf x}\tau)]
F_{\alpha\beta}^{\dagger}({\bf x}\tau,{\bf x}^{\prime}\tau^{\prime})
-U_{\gamma\alpha}({\bf x})
F_{\gamma\beta}^{\dagger}({\bf x}\tau,{\bf x}^{\prime}\tau^{\prime})& 
\nonumber \\
&-\int d{\bf x}^{\prime\prime}
\Delta^{\dagger}_{\alpha\gamma}({\bf x}\tau^{0+},{\bf x}^{\prime\prime}\tau)
G_{\gamma\beta}({\bf x}^{\prime\prime}\tau,{\bf x}^{\prime}\tau^{\prime})
=0\;.& 
\end{eqnarray}
\end{mathletters}
Here repeated spin indices mean summing over these indices. In these 
equations, 
\begin{equation}
h({\bf x}\tau)=\frac{[{\bf p}+e{\bf A}({\bf 
x}\tau)]^{2}}{2m}-e\varphi({\bf x}\tau)-\mu\;,
\end{equation}
is the single-electron ($-e$) Hamiltonian 
with ${\bf A}({\bf x}\tau)$, $\varphi ({\bf x}\tau)$, and $\mu$
denoting the vector, scalar, and chemical potentials. (We have assumed
$\hbar = c = 1$.)
By assuming zero-range interactions bewteen electrons 
and impurities, the impurity scattering potential can be written as  
\begin{equation}
U_{\alpha\beta}({\bf x})
=\sum_{i\in I}[U_{1}\delta_{\alpha\beta}+U_{2}({\bf S}_{i}\cdot 
\frac{\mbox{\boldmath{$\sigma$}}_{\alpha\beta}}{2})]
\delta({\bf x}-{\bf R}_{i})\;, 
\end{equation}
where $I$ denotes the set of impurity 
sites, $\mbox{\boldmath{$\sigma$}}$ is made of the Pauli spin matrices, 
${\bf S}_{i}$ is the spin carried by an impurity at ${\bf R}_{i}$.
$U_{1}$ and $U_{2}$ are the non-spin-flip and spin-flip interaction 
strengths, respectively.
By definition, the order parameter in real coordinate and imaginary 
time space is 
\begin{equation}
\label{EQ:Gap0}
\Delta^{*}_{\alpha\beta}({\bf x}\tau,{\bf x}^{\prime}\tau)
=V({\bf x}-{\bf x}^{\prime}) 
F_{\alpha\beta}^{\dagger}({\bf x}\tau^{0+},{\bf x}^{\prime}\tau)\;, 
\end{equation}
where $-V({\bf x}-{\bf x}^{\prime})$ is the effective pairing interaction 
between electrons. 
Because of the spatial and temporal non-uniformity,
the Green function $G_{\alpha\beta}({\bf x}\tau,{\bf
x}^{\prime}\tau^{\prime})$ and
$F^{\dagger}_{\alpha\beta}({\bf x}\tau,
{\bf x}^{\prime}\tau^{\prime})$ are not the
functions of coordinate and time differences. When expressed in the
imaginary frequency space after the Fourier transform, they depend
on two frequency variables. For the spatial coordinate dependence, 
as treated in the static case,~\cite{RXT95} we
express these two functions in terms of the center-of-mass 
coordinate ${\bf R} = ({\bf x} + {\bf x'})/2$ and the relative momentum 
after a Fourier transform with respect to the relative coordinate 
${\bf r} ={\bf x} - {\bf x'}$.
Thus  Eq.~(\ref{EQ:Gap0}) can be rewritten as:
\begin{equation}
\label{EQ:Gap}
\Delta_{\alpha\beta}^{*}({\bf R},{\bf k};\omega)
=T\sum_{\epsilon}\int \frac{d{\bf k}^{\prime}}{(2\pi)^{2}}
V({\bf k}-{\bf k}^{\prime})
F^{\dagger}_{\alpha\beta}({\bf R},{\bf 
k}^{\prime};\epsilon,\epsilon-\omega)\;, 
\end{equation}    
where  
${\bf k}$ is the relative momentum,
$\omega=2in\pi T$ and $\epsilon=i(2n^{\prime}+1)\pi T$ 
with integers $n$ and $n^{\prime}$, 
$-V({\bf k}-{\bf k}^{\prime})$ is the pairing interaction in the momentum 
space, $F^{\dagger}({\bf R},{\bf k};\epsilon,\epsilon^{\prime})$ is the 
Fourier transform of 
$F^{\dagger}({\bf x}\tau,{\bf x}^{\prime}\tau^{\prime})$.
To relate to high-$T_{c}$ superconductors, we have assumed that the 
system under consideration is two dimensional. 
For the spin-singlet pairing, the order parameter is given 
in the spin space as $\Delta^{*}_{\alpha\beta}
=\Delta^{*}g_{\alpha \beta}$, where 
\begin{equation}
g_{\alpha\beta}=\left( \begin{array}{cc}
0 & 1 \\
-1 &  0 
\end{array} \right)_{\alpha\beta} \;. 
\end{equation}
To obtain the TDGL equations for superconductors 
of a mixed $d$- and $s$-wave symmetry, 
we make the following ansatz for 
the pairing interaction and the order parameter
\begin{equation}
V({\bf k}-{\bf k}^{\prime})=V_{s}+V_{d}(\hat{k}_{x}^{2}-\hat{k}_{y}^{2}) 
(\hat{k}_{x}^{\prime 2}-\hat{k}_{y}^{\prime 2})\;,
\end{equation}
\begin{equation}
\Delta^{*}({\bf R},{\bf k};\omega)=\Delta^{*}_{s}({\bf R};\omega)
+\Delta^{*}_{d}({\bf R};\omega)(\hat{k}_{x}^{2}-\hat{k}_{y}^{2})\;,
\end{equation}
where $V_{d}$ and $V_{s}$ are positive so that both the $d$- and 
$s$-channel interactions are attractive.  
The $d$-channel 
attractive interaction could originate from the antiferromagnetic 
spin fluctuation, whereas the $s$-channel attractive 
interaction might arise from phonon mediation. 

Introducing the Green function $\tilde{G}^{0}$ of the normal metal, which 
satisfies the equation 
\begin{equation}
[-\frac{\partial}{\partial \tau}-h({\bf x}\tau) ] 
\tilde{G}_{\alpha\beta}^{0}({\bf x}\tau, {\bf x}^{\prime}\tau^{\prime})
-U_{\alpha\gamma}({\bf x})
\tilde{G}_{\gamma\beta}^{0}({\bf x}\tau,{\bf 
x}^{\prime}\tau^{\prime})
=\delta({\bf x}-{\bf x}^{\prime})\delta(\tau-\tau^{\prime})
\delta_{\alpha\beta}\;,
\end{equation}
Eq.~(\ref{EQ:motion}) may be converted to a 
set of coupled integral equations:
\begin{eqnarray}
G_{\alpha\beta}({\bf x}\tau,{\bf x}^{\prime}\tau^{\prime})
&=&\tilde{G}^{0}_{\alpha\beta}({\bf x}\tau,{\bf x}^{\prime}\tau^{\prime})
\nonumber \\
&&-\int d{\bf x}_{1} d{\bf x}_{2}  
d\tau_{1}  \tilde{G}^{0}_{\alpha\mu}({\bf x}\tau,{\bf x}_{1}\tau_{1})
\Delta_{\mu\nu}({\bf x}_{1}\tau_{1},{\bf x}_{2}\tau_{1})
F^{\dagger}_{\nu\beta}({\bf x}_{2}\tau_{1},
{\bf x}^{\prime}\tau^{\prime})\;, \label{EQ:green1}
\end{eqnarray}
\begin{equation}
\label{EQ:green2}
F^{\dagger}_{\alpha\beta}({\bf x}\tau,{\bf x}^{\prime}\tau^{\prime})
=\int d{\bf x}_{1}d{\bf x}_{2}d\tau_{1} 
\tilde{G}^{0}_{\mu\alpha}({\bf x}_{1}\tau_{1},{\bf x}\tau) 
\Delta^{*}_{\mu\nu}({\bf x}_{1}\tau_{1},{\bf x}_{2}\tau_{1}) 
G_{\nu\beta}({\bf x}_{2}\tau_{1},{\bf x}^{\prime}\tau^{\prime})\;.
\end{equation}
Also note that the normal-state Green function can in turn be 
written as an integral equation
\begin{eqnarray}
\tilde{G}^{0}_{\alpha\beta}({\bf x}\tau,{\bf x}^{\prime}\tau^{\prime})
&=&G^{0}_{\alpha\beta}({\bf x}\tau,{\bf x}^{\prime}\tau^{\prime})
+\int d{\bf x}^{\prime\prime}d\tau^{\prime\prime} 
G^{0}_{\alpha\gamma}({\bf x}\tau,{\bf x}^{\prime\prime}\tau^{\prime\prime})
\nonumber \\
&& \times 
[\frac{e{\bf A}({\bf x}^{\prime\prime}\tau^{\prime\prime})\cdot 
{\bf p}_{{\bf x}^{\prime\prime}}}{m}
-e\varphi({\bf x}^{\prime\prime}\tau^{\prime\prime})]
\tilde{G}^{0}_{\gamma\beta}
({\bf x}^{\prime\prime}\tau^{\prime\prime},{\bf x}^{\prime}\tau^{\prime})  
\;,
\end{eqnarray}
with  $G^{0}$ as the normal-state single-particle Green function 
in the absence of the electromagnetic field but including the effect due to
impurity scatterings. To write down the above integral equation, the 
squared term of the vector potential has been neglected and the Coulomb 
gauge is chosen. 

\subsection{Analytical continuation}

To incorporate the time dependence of physical quantities, we use the 
analytical continuation technique 
discussed in Refs.~\cite{Gor68,Eliash69} to transform 
imaginary frequencies into real frequencies.
The procedure is as follows:
(i) In Eq.~(\ref{EQ:Gap}), each term of the summation over the imaginary
frequency $\epsilon$ can be regarded as the residue of an integral
along the contour around the point $z=\epsilon$ so that we have the
transformation $T\sum_{\epsilon}\rightarrow \frac{1}{4\pi i}\oint_{\cal C}
dz \tanh \frac{z}{2T}$. Associated with this transformation, all involved
$\epsilon$  are replaced with $z$. For example, 
\begin{equation}
\label{EQ:Contour1}
T\sum_{\epsilon}
G^{0}(-\epsilon)G^{0}(\epsilon-\omega^{\prime})=\frac{1}{4\pi i}
\oint_{\cal C} dz \tanh \frac{z}{2T} G^{0}(-z)G^{0}(z-\omega^{\prime})\;,
\end{equation}
where the spatial and spin variables have been 
suppressed for simplicity.
(ii) Deform the contour integral around $z$ into the straight line integrals
along $z=\epsilon\pm i0^{+}$, $z=\epsilon+\omega^{\prime} \pm
i0^{+}$, \dots,
where $\epsilon\in (-\infty,\infty)$ is the real integral variable and
  $0^{+}$ is infinitesimal.
So far, $\omega^{\prime}$, \dots are still imaginary frequency $2n\pi iT$
and we take $n\ge 0$ since we will perform the analytical continuation from
the upper half-plane.
As a consequence, each Green function $G^{0}(z)$ with energy variable
coincident with this line is decomposed into
$G^{0(R)}(\epsilon)-G^{0(A)}(\epsilon)$,
where $G^{0(R,A)}$
are the retarded and advanced Green functions, respectively.
The minus sign before $G^{0(A)}$ comes from changing the direction of
integration.
The other Green functions are mapped
to the retarded or advanced Green function depending on their energy
variable. Then Eq.~(\ref{EQ:Contour1}) becomes
\begin{eqnarray}
&\frac{1}{4\pi i}\int_{-\infty}^{\infty} d\epsilon
\tanh \frac{\epsilon}{2T} \{[G^{0(R)}(-\epsilon)-G^{0(A)}(-\epsilon)]
G^{0(A)}(\epsilon-\omega^{\prime}) &\nonumber \\
&+G^{0(R)}(-(\epsilon+\omega^{\prime}))
[G^{0(R)}(\epsilon)-G^{0(A)}(\epsilon)]\}\;.&\label{EQ:Contour2}
\end{eqnarray}
(iii) Since all $\omega^{\prime}$, \dots, lie in the upper half-plane,
the Fourier transform of the expansions~(\ref{EQ:green1}) and
(\ref{EQ:green2}) are
analytical in each of these variables.
Therefore, we can implement the analytical continuation by simply
replacing all $\omega^{i}$ with $\omega^{i}+i0^{+}$.
Simultaneously the discrete 
summation $T\sum_{\omega^{\prime}}$ is
replaced by a continuous integral 
$(2\pi)^{-1}\int d\omega^{\prime}$ and
the Kronecker delta $T^{-1}\delta_{\omega,\omega^{\prime}}$ by
the Dirac delta $(2\pi)\delta(\omega-\omega^{\prime})$.
Finally, we get
\begin{eqnarray}
&\frac{1}{4\pi i} \int_{-\infty}^{\infty} d\epsilon
[-\tanh \frac{\epsilon}{2T}
G^{0(A)}(-\epsilon)G^{0(A)}(\epsilon-\omega^{\prime})
+\tanh \frac{\epsilon-\omega^{\prime}}{2T}G^{0(R)}(-\epsilon)
G^{0(R)}(\epsilon-\omega^{\prime})]& \nonumber \\
&+\frac{1}{4\pi i} \int_{-\infty}^{\infty} d\epsilon
G^{0(R)}(-\epsilon)
[\tanh \frac{\epsilon}{2T}-\tanh\frac{\epsilon-\omega^{\prime}}{2T}]
G^{0(A)}(\epsilon-\omega^{\prime})\;.&
\end{eqnarray}  
The first two terms consist of the only advanced and the only retarded
Green functions (Following Ref.~\cite{Gor68}, 
we shall refer to them as the normal part). The rest part has those 
terms involving the product of retarded and advanced Green functions, in 
which a change from a retarded to an advanced Green function occurs 
in only one place (We shall refer to them as the anomalous parts).
After obtaining the results in terms of real frequency,
we can  perform the inverse Fourier transform to represent them in a real
time.

Using these rules, we can obtain the following expression 
\begin{eqnarray}
T\sum_{\epsilon}F^{\dagger}(\epsilon,\epsilon-\omega)& \rightarrow & 
\frac{1}{4\pi i} \int d\epsilon \tanh \frac{\epsilon}{2T} 
[F^{\dagger(R)}(\epsilon+\omega,\epsilon)
-F^{\dagger(A)}(\epsilon,\epsilon-\omega)] \nonumber \\
&&-\frac{1}{4\pi i} \int d\epsilon \int d\epsilon_{1} \{
[G^{+(R)}(\epsilon,\epsilon_{1})\Delta_{\omega^{\prime}}^{*}
G^{-(A)}(\epsilon_{1}-\omega^{\prime},\epsilon-\omega)
\nonumber \\
&&
+ F^{\dagger(R)}(\epsilon,\epsilon_{1})\Delta_{\omega^{\prime}}
F^{(A)}(\epsilon_{1}-\omega^{\prime},\epsilon-\omega)]
\nonumber \\
&&+[G^{+(R)}(\epsilon,\epsilon_{1})
F^{(A)}(\epsilon_{1}-\omega^{\prime},\epsilon-\omega)
+ F^{\dagger(R)}(\epsilon,\epsilon_{1})
G^{-(A)}(\epsilon_{1}-\omega^{\prime},\epsilon-\omega)]
\nonumber \\
&&\times 
[e{\bf A}_{\omega^{\prime}}\cdot {\bf p}/m-e\varphi_{\omega^{\prime}}]\}
[\tanh\frac{\epsilon_{1}}{2T}-
\tanh\frac{\epsilon_{1}-\omega^{\prime}}{2T}]\;,
\end{eqnarray}
where, 
$G^{-(R,A)}$ and $F^{\dagger(R,A)}$ are formally defined by 
\begin{eqnarray}
G^{-}({\bf p},{\bf p}-{\bf k};\epsilon,\epsilon-\omega)&=& 
G^{0-}({\bf p};\epsilon)
+G^{0-}({\bf p};\epsilon)\Delta_{\omega^{\prime}}({\bf k}^{\prime})
G^{0-}({\bf p}-{\bf k}^{\prime};\epsilon-\omega^{\prime})  
\Delta_{\omega^{\prime\prime}}^{*}({\bf k}^{\prime\prime})\nonumber \\
&&\times 
G^{0-}({\bf p}-{\bf k}^{\prime}-{\bf k}^{\prime\prime}
;\epsilon-\omega^{\prime}-\omega^{\prime\prime}) +\dots\;,
\end{eqnarray} 
and 
\begin{eqnarray}
F^{\dagger}({\bf p},{\bf p}-{\bf k};\epsilon,\epsilon-\omega)&=& 
-\{ G^{0+}({\bf p};\epsilon)\Delta_{\omega^{\prime}}^{*}({\bf k}^{\prime})
G^{0-}({\bf p}-{\bf k}^{\prime};\epsilon-\omega^{\prime})+\dots \} \;,
\end{eqnarray} 
in which the substitution 
$\epsilon \rightarrow \epsilon \pm i 0^{+}$ for the retarded (advanced) 
Green function should be made. Here 
$G^{0\pm}({\bf p};\epsilon)=[\epsilon\pm \xi_{\bf p}]^{-1}$ and all 
$\omega^{i}$ are real.  The functions $G^{+}$ and $F$  are obtained from 
$G^{-}$ and $F^{\dagger}$ by changing 
the sign of $\xi$ in $G^{0\pm}$. Note that $G^{+}$ and $F$ are introduced 
only for simplicity of notation. 

\subsection{Normal part}
The normal part of $\Delta^{*}$ can be written as 
\begin{equation}
\label{EQ:Gap-norm}
\Delta_{\alpha\beta}^{*,N}({\bf R},{\bf k};\omega)
=\int \frac{d{\bf k}^{\prime}}{(2\pi)^{2}}
V({\bf k}-{\bf k}^{\prime})
[T\sum_{\epsilon_{n}\geq 0}F^{\dagger(R)}_{\alpha\beta}({\bf R},{\bf 
k}^{\prime};\epsilon_{n}+\omega,\epsilon_{n})
+T\sum_{\epsilon_{n}\leq 0}F^{\dagger(A)}_{\alpha\beta}({\bf R},{\bf 
k}^{\prime};\epsilon_{n},\epsilon_{n}-\omega)]
\;. 
\end{equation}    

The evaluation of the normal part can be done  by expanding the 
expressions in powers 
of the order parameter. We write for $F^{\dagger(R,A)}$ and $G^{(R,A)}$ 
up to terms of the third and second order in $\Delta$, respectively,  
so that 
\begin{eqnarray}
F_{\alpha\beta}^{\dagger(R,A)}({\bf x},{\bf 
x}^{\prime};\epsilon,\epsilon-\omega)
&=&F_{I1,\alpha\beta}^{\dagger(R,A)}({\bf x},{\bf x}^{\prime};
\epsilon,\epsilon-\omega)
+F_{I2,\alpha\beta}^{\dagger(R,A)}({\bf x},{\bf x}^{\prime};
\epsilon,\epsilon-\omega)
\nonumber \\
&&+F_{II,\alpha\beta}^{\dagger(R,A)}({\bf x},{\bf x}^{\prime};
\epsilon,\epsilon-\omega)\;,
\end{eqnarray}
where
\begin{eqnarray}
F^{\dagger(R,A)}_{I1,\alpha\beta}({\bf x},{\bf x}^{\prime};
\epsilon,\epsilon-\omega)&=&\int d{\bf x}_{1} d{\bf x}_{2}
G^{0(R,A)}_{\mu\alpha}({\bf x}_{1},{\bf x};-\epsilon)
e^{-ie{\bf A}_{\omega^{\prime}}({\bf x}_{1})\cdot ({\bf x}_{1}-{\bf x})}
\Delta^{*}_{\mu\nu}({\bf x}_{1},{\bf x}_{2};\omega^{\prime\prime})
\nonumber \\
&& \times
G^{0(R,A)}_{\nu\beta}({\bf x}_{2},{\bf x}^{\prime};\epsilon-\omega)
e^{-ie{\bf A}_{\omega^{\prime\prime\prime}}({\bf x}_{2})\cdot
({\bf x}_{2}-{\bf x}^{\prime})}\;,\label{EQ:anorm1}
\end{eqnarray}
\begin{eqnarray}
F^{\dagger(R,A)}_{I2,\alpha\beta}({\bf x},{\bf
x}^{\prime};\epsilon,\epsilon-\omega)
&=&\int d{\bf x}_{1} d{\bf x}_{2} d{\bf x}_{3}
G^{0(R,A)}_{\mu\alpha}({\bf x}_{1},{\bf x};-\epsilon)
\Delta^{*}_{\mu\nu}
({\bf x}_{1},{\bf x}_{2};\omega^{\prime})
G^{0(R,A)}_{\nu\rho}({\bf x}_{2},{\bf x}_{3};\epsilon-\omega^{\prime})
\nonumber \\
&& \times
[-e\varphi_{\omega^{\prime\prime}}({\bf x}_{3})]
G^{0(R,A)}_{\rho\beta}({\bf
x}_{3},{\bf
x}^{\prime}; \epsilon-\omega) \nonumber \\
&&+\int d{\bf x}_{1}d{\bf x}_{2}d{\bf x}_{3}  
G^{0(R,A)}_{\mu\rho}({\bf x}_{1},{\bf x}_{3};-(\epsilon-\omega^{\prime}))
[-e\varphi_{\omega^{\prime}}({\bf x}_{3})]
G^{0(R,A)}_{\rho\alpha}({\bf x}_{3},{\bf x};  
-\epsilon)
\nonumber \\
&& \times
\Delta^{*}_{\mu\nu}({\bf x}_{1},{\bf
x}_{2};\omega^{\prime\prime}) 
G^{0(R,A)}_{\nu\beta}({\bf x}_{2},{\bf x}^{\prime};
\epsilon-\omega)\;,\label{EQ:anorm2}
\end{eqnarray}
\begin{eqnarray}
F^{\dagger(R,A)}_{II,\alpha\beta}({\bf x},{\bf 
x}^{\prime};\epsilon,\epsilon-\omega)    
&=&-\int d{\bf x}_{1}d{\bf x}_{2} d{\bf x}_{3} d{\bf x}_{4} d{\bf x}_{5}
d{\bf x}_{6} G^{0(R,A)}_{\mu\alpha}({\bf x}_{1},{\bf x};-\epsilon)
\Delta^{*}_{\mu\nu}({\bf x}_{1},{\bf x}_{2};\omega^{\prime})
\nonumber \\
&& \times
G^{0(R,A)}_{\nu\lambda}({\bf x}_{2},{\bf x}_{3};\epsilon-\omega^{\prime})
\Delta_{\lambda\sigma}({\bf x}_{3},{\bf x}_{4};\omega^{\prime\prime})
G^{0(R,A)}_{\rho\sigma}({\bf x}_{5},{\bf x}_{4};
-(\epsilon-\omega^{\prime}-\omega^{\prime\prime}))
\nonumber \\
&& \times
\Delta^{*}_{\rho\tau}({\bf x}_{5},{\bf x}_{6};\omega^{\prime\prime\prime})
G^{0(R,A)}_{\tau\beta}({\bf x}_{6},{\bf x}^{\prime};\epsilon-\omega)
\;.\label{EQ:anorm3}
\end{eqnarray}
Here the summation over the imaginary frequency $\omega^{i}$ with   
the constraint $\sum_{i} \omega^{i}=\omega$ is implied.
To write down the above expression for $F^{\dagger(R,A)}$, we have expanded 
$\tilde{G}^{0}$ to the first order in the scalar potential
$-e\varphi$, and separate this expansion term out explicitly. 
As far as the dependence of $\tilde{G}^{0}$ on the magnetic field is 
concerned, the quasiclassical phase approximation can be used to write it in 
the form
$G^{0}({\bf x}, {\bf x}^{\prime};\epsilon)
\exp[-ie{\bf A}_{\omega} \cdot ({\bf x}-{\bf x}^{\prime})]$.   
Accordingly, the gap function can also be written as
a sum of three parts
\begin{equation}
\label{EQ:gap-norm1}
\Delta_{\alpha\beta}^{*,N}({\bf R},{\bf k};\omega)
=\Delta_{I1,\alpha\beta}^{*,N}({\bf R},{\bf k};\omega)
+\Delta_{I2,\alpha\beta}^{*,N}({\bf R},{\bf k};\omega)
+\Delta_{II,\alpha\beta}^{*,N}({\bf R},{\bf k};\omega)\;.
\end{equation}

The remaining task involves the evaluation of the average over an ensemble
of randomly distributed configurations. 
As an approximation,
$\Delta^{*}$ is regarded as very nearly independent of impurity 
configurations. 
%We are now   
%encountered with the evaluation of the average of the product of two
%Green functions $\langle G^{0}G^{0}\rangle$.
%Using the impurity averaging
%diagram technique, the averaged normal-state single-particle Green
%function in the absence of the external field can be easily found
%\begin{equation}
%G_{\alpha\beta}^{0(R,A)}({\bf k},\epsilon_{n})=
%\frac{\delta_{\alpha\beta}}{\epsilon_{n}
%-\xi_{\bf k}\pm \frac{i}{2\tau_{1}}}\;,
%\end{equation}
%where $\xi_{\bf k}$ is the kinetic energy 
We assume that impurities density $n_{i}$
are randomly distributed and their spins are
arbitrarily oriented so that there is no correlation among 
them.~\cite{AG60} 
Using the Born approximation, we can show that the impurity averaged 
zero-field normal-state
Green function takes the following form
\begin{equation}
\langle G_{\alpha\beta}^{0}({\bf x},{\bf x}^{\prime};\epsilon_n)
\rangle
={1\over (2\pi)^2}
\int  d{\bf k} e^{i{\bf k}({\bf x}-{\bf x}^{\prime})}
{\delta_{\alpha\beta}\over i\epsilon_{n}\eta_{1}-\xi_{\bf k}},
\end{equation}
where  $\langle \dots \rangle$ denotes the average over
the impurity configuration, $\xi_{\bf k}=\frac{{\bf k}^{2}}{2m}-\mu$ 
is the 
kinetic energy, and $\eta_{1}=1+({2\tau_{1}|\epsilon_{n}|})^{-1}$ with
the scattering time $\tau_{1}$ given by
\begin{equation}
\frac{1}{\tau_{1}}=2\pi n_{i}N(0) [\vert U_{1} \vert^{2}
+\frac{1}{4}S(S+1) \vert U_{2} \vert^{2}]\;.
\end{equation}  
Here $n_{i}$ is the impurity density and $N(0)$ is the density of states
at the Fermi surface per spin.
The evaluation of the product of Green functions can be 
conveniently performed based on the diagrammatic rule.~\cite{XRT96,KT97}
If there are two Green functions connected by an $s$-wave order parameter or
$s$-channel two-body interaction, these two Green functions
might be called directly connected and  we should attach a vertex
renormalization factor: 
\begin{eqnarray}
\eta^{(R,A)}(\epsilon) &=& [1-a^{(R,A)}(0)]^{-1} \nonumber \\
&\approx & \frac{\epsilon\pm \frac{i}{2\tau_{1}}}
{\epsilon\pm \frac{i}{\tau_{s}}}\mp
\frac{\frac{i\omega}{\tau_{2}}}{(2\epsilon\pm
\frac{2i}{\tau_{s}})^{2}}\;,
\label{EQ:renorm1}
\end{eqnarray}
where
\begin{equation}
a^{(R)}(0)=\frac{1}{2\pi N(0)\tau_{2}}\int \frac{d{\bf p}}{(2\pi)^{2}} 
G^{0(R)}({\bf p},-(\epsilon-i\omega))
G^{0(R)}({\bf p},\epsilon)\;,
\end{equation}
and 
\begin{equation}
a^{(A)}(0)=\frac{1}{2\pi N(0)\tau_{2}}\int \frac{d{\bf p}}{(2\pi)^{2}} 
G^{0(A)}({\bf p},-\epsilon)
G^{0(A)}({\bf p},\epsilon+i\omega)\;.
\end{equation}
Here 
\begin{equation}
\frac{1}{\tau_{2}}=2\pi n_{i}N(0) [\vert U_{1}\vert^{2}
-\frac{1}{4}S(S+1)\vert U_{2} \vert^{2}] \;,
\end{equation}
and the spin-flip scattering rate is defined as
$2\tau_{s}^{-1}=\tau_{1}^{-1}-\tau_{2}^{-1}$. 
If the two Green functions are connected by
a $d$-wave order parameter or $d$-channel two-body interaction, 
they might be called not directly connected and
we have no vertex correction. For the average of the product of more
than two Green functions, an impurity line can also appear across the box
of a diagram. Because of their zero
contribution, diagrams with more than one impurity line across the box 
should not be included.  
This impurity averaging technique was first used by 
Abrikosov and Gorkov~\cite{AG60} for conventional $s$-wave superconductors.
Recent experimental measurements by Bernhard {\em et al.}~\cite{BTBR96} 
on various types of YBa$_{2}$(Cu$_{1-x}$Zn$_{x}$)$_{3}$O$_{7-\delta}$ 
samples have shown that the depression of $T_{c}$ by Zn doping can be fitted 
well with the Abrikosov-Gorkov theory applied to the $d$-wave 
superconductivity.  

Now we give a derivation for the gap function 
from $T\sum_{\epsilon_{n}\geq 
0}F^{\dagger(R)}({\epsilon_{n}+\omega,\epsilon_{n}})$. 
The contribution from 
$T\sum_{\epsilon_{n}\leq 0}F^{\dagger(R)}
({\epsilon_{n},\epsilon_{n}-\omega})$
can be obtained by merely changing all explicit $i$ to $-i$ and $\omega$ to 
$-\omega$, which gives the same result.  
In addition, one can easily see that 
$\Delta_{I2,\alpha\beta}^{*,N}({\bf R},{\bf k};\omega)=0$
since the contribution from the two terms 
given by Eq.~(\ref{EQ:anorm2}) cancelled with each other.  
Therefore, we obtain  
\begin{eqnarray}
\Delta_{I1,\alpha\beta}^{*,N(R)}({\bf R},{\bf k};\omega)&=&
T\sum_{\epsilon_{n}\geq 0}\int \frac{d{\bf k^{\prime}}}{(2\pi)^{2}} 
V({\bf k}-{\bf k}^{\prime}) \int d{\bf r}e^{-i{\bf k}^{\prime}\cdot 
{\bf r}} \int \int d{\bf R}^{\prime} d{\bf r}^{\prime} 
\nonumber \\
&&\times 
\langle G^{0(R)}_{\lambda \alpha}
({\bf R}^{\prime}+\frac{\bf r^{\prime}}{2}, {\bf R}+\frac{\bf 
r}{2};-(\epsilon_{n}+\omega))g_{\lambda\mu}G^{0(R)}_{\mu\beta}
( {\bf R}^{\prime}-\frac{\bf r^{\prime}}{2}, {\bf R}-\frac{\bf 
r}{2};\epsilon_{n})\rangle \nonumber \\
&&\times e^{i({\bf R}^{\prime}-{\bf R})\cdot {\bf 
\Pi} }\int \frac{d {\bf k}^{\prime\prime} }{(2\pi)^{2}}
e^{ i{\bf k}^{\prime\prime}\cdot {\bf r}^{\prime} }
\Delta_{\omega}^{*}({\bf R}, {\bf k}^{\prime\prime})\;,
\end{eqnarray}
and 
\begin{eqnarray}
\Delta_{II,\alpha\beta}^{*,N(R)}({\bf R},{\bf k};\omega)
&=&T\sum_{\epsilon_n\geq 0}\int 
\frac{d{\bf k}^{\prime}}{(2\pi)^{2}}V({\bf k}-{\bf k}^{\prime}) 
\int d{\bf r} e^{-i{\bf k}^{\prime} \cdot {\bf r}}
\int d{\bf R}^{\prime} d{\bf r}^{\prime} 
d{\bf R}_{1}d{\bf r}_{1}
d{\bf R}_{2}d{\bf r}_{2}
\nonumber\\
&&\times
\langle
G_{\lambda\alpha}^{0(R)}
({\bf R}^{\prime}+{{\bf r}^{\prime}\over2},{\bf R}+{{\bf 
r}\over2},-\epsilon_{n}) g_{\lambda\mu}
{G}_{\mu\nu}^{0(R)}
({\bf R}^{\prime}-{{\bf r}^{\prime}\over2},{\bf R}_{1}+{{\bf 
r}_{1}\over2},\epsilon_{n}) \nonumber\\
&&\times g_{\nu\rho}
{G}_{\kappa\rho}^{0(R)}
({\bf R}_{2}+{{\bf r}_{2}\over2},{\bf R}_{1}-{{\bf r}_{1}\over2},
-\epsilon_{n})
g_{\kappa\sigma}
{G}_{\sigma\beta}^{0(R)}  
({\bf R}_{2}-{{\bf r}_{2}\over2},{\bf R}-{{\bf r}\over2},\epsilon_{n})
\rangle
\nonumber\\  
&&\times\int{d{\bf k}^{\prime\prime}d{\bf k}_{1}d{\bf k}_{2}\over (2\pi)^6}
\Delta^{*}_{\omega^{\prime}}({\bf R},{\bf k}^{\prime})
\Delta_{\omega^{\prime\prime}} ({\bf R},{\bf k}_{1})
\Delta^{*}_{\omega^{\prime\prime\prime}}({\bf R},{\bf k}_{2})
e^{i({\bf k}^{\prime\prime}\cdot{\bf r}^{\prime}+{\bf k}_{1}\cdot{\bf r}_{1}
+{\bf k}_{1}\cdot{\bf r}_{2})}\;.
\end{eqnarray}
Here ${\bf \Pi}=-i\nabla_{\bf R}-2e{\bf A}_{\omega^{\prime}}({\bf R})$ 
and we have assumed the slow variation of the vector 
potential. 

Since $G^{0}_{\alpha\beta}=G^{0}\delta_{\alpha\beta}$ is diagonal in the 
spin space, we can express $\Delta_{I1,\alpha\beta}^{*,N(R)}
=\Delta_{I1}^{*,N(R)}g_{\alpha\beta}$ with 
\begin{equation}
\Delta_{I1}^{*,N(R)}({\bf R},{\bf k};\omega)=
\Delta_{I1c}^{*,N(R)}({\bf R},{\bf k};\omega)+
\Delta_{I1g}^{*,N(R)}({\bf R},{\bf k};\omega)\; ,
\end{equation} 
where 
\begin{eqnarray}
\Delta_{I1c}^{*,N(R)}({\bf R},{\bf k})&=&
T\sum_{\epsilon_{n}\geq 0}\int \frac{d{\bf k^{\prime}}}{(2\pi)^{2}} 
V({\bf k}-{\bf k}^{\prime}) \int d{\bf r}e^{-i{\bf k}^{\prime}\cdot 
{\bf r}} \int \int d{\bf R}^{\prime} d{\bf r}^{\prime} 
\nonumber \\
&&\times 
\langle G^{0(R)}_{\lambda \alpha}
({\bf R}^{\prime}+\frac{\bf r^{\prime}}{2}, {\bf R}+\frac{\bf 
r}{2};-(\epsilon_{n}+\omega))g_{\lambda\mu}G^{0(R)}_{\mu\beta}
( {\bf R}^{\prime}-\frac{\bf r^{\prime}}{2}, {\bf R}-\frac{\bf 
r}{2};\epsilon_{n})\rangle \nonumber \\
&&\times \int \frac{d {\bf k}^{\prime\prime} }{(2\pi)^{2}}
e^{ i{\bf k}^{\prime\prime}\cdot {\bf r}^{\prime} }
\Delta_{\omega}^{*}({\bf R}, {\bf k}^{\prime\prime})\;,
\end{eqnarray}
and
\begin{eqnarray}
\Delta_{Ig}^{*}({\bf R},{\bf k})&=&-{1\over2}
T\sum_{\epsilon_{n}\geq 0}\int \frac{d{\bf k^{\prime}}}{(2\pi)^{2}} 
V({\bf k}-{\bf k}^{\prime}) \int d{\bf r}e^{-i{\bf k}^{\prime}\cdot 
{\bf r}} \int \int d{\bf R}^{\prime} d{\bf r}^{\prime} 
\nonumber \\
&&\times 
\langle G^{0(R)}_{\lambda \alpha}
({\bf R}^{\prime}+\frac{\bf r^{\prime}}{2}, {\bf R}+\frac{\bf 
r}{2};-(\epsilon_{n}+\omega))g_{\lambda\mu}G^{0(R)}_{\mu\beta}
( {\bf R}^{\prime}-\frac{\bf r^{\prime}}{2}, {\bf R}-\frac{\bf 
r}{2};\epsilon_{n})\rangle \nonumber \\
&&\times [({\bf R}^{\prime}-{\bf R})\cdot {\bf 
\Pi}]^{2} \int \frac{d {\bf k}^{\prime\prime} }{(2\pi)^{2}}
e^{ i{\bf k}^{\prime\prime}\cdot {\bf r}^{\prime} }
\Delta_{\omega}^{*}({\bf R}, {\bf k}^{\prime\prime})\;,
\end{eqnarray}
Similarly, 
$\Delta_{II,\alpha\beta}^{*,N(R)}
=\Delta_{II}^{*,N(R)}g_{\alpha\beta}$.

Using the diagrammatic rule mentioned above, we calculate 
$\Delta_{Ic}^{*,N(R)}$, which in general has four terms for the mixed 
$s$ and $d$ wave superconductors, 
$\Delta_{Ic}^{*,N(R)}=\sum_{i=1}^{4} \Delta_{Ic,i}^{*,N(R)}$.
It is easy to show that 
\begin{mathletters}
\label{EQ:OD-C}
\begin{eqnarray}
\Delta_{Ic,1}^{*,N(R)}({\bf R}, {\bf k};\omega)
&=&T\sum_{\epsilon_{n}\geq 0}\int \frac{d{\bf p}}{(2\pi)^{2}} 
\eta^{(R)}V_{s}G^{0(R)}(-{\bf p};-(\epsilon_{n}+\omega))
G^{0(R)}({\bf p}; \epsilon_{n})\Delta_{s}^{*}({\bf R},\omega)
\nonumber \\
&=&\frac{V_{s}N(0)}{2}
\{ [\ln\frac{2e^{\gamma}\omega_{D}}{\pi T}+\psi(\frac{1}{2})-
\psi(\frac{1}{2}+\frac{\rho_{s}}{2})] 
+\frac{i\omega}{4\pi T}\psi^{\prime}(\frac{1}{2}+\frac{\rho_{s}}{2})\}
\nonumber \\
&&\times \Delta_{s}^{*}({\bf R},\omega)\;,
\end{eqnarray}

\begin{eqnarray}
\Delta_{Ic,2}^{*,N(R)}({\bf R}, {\bf k};\omega)
&=&T\sum_{\epsilon_{n}\geq 0}\int \frac{d{\bf p}}{(2\pi)^{2}} 
\eta^{(R)}V_{s}G^{0(R)}(-{\bf p};-(\epsilon_{n}+\omega))
G^{0(R)}({\bf p}; \epsilon_{n})\Delta_{d}^{*}({\bf R},\omega)
(\hat{p}_{x}^{2}-\hat{p}_{y}^{2}) \nonumber \\
&=&0\;,
\end{eqnarray}

\begin{eqnarray}
\Delta_{Ic,3}^{*,N(R)}({\bf R}, {\bf k};\omega)
&=&T\sum_{\epsilon_{n}\geq 0}\int \frac{d{\bf p}}{(2\pi)^{2}} 
\eta^{(R)} V_{d}
(\hat{k}_{x}^{2}-\hat{k}_{y}^{2})(\hat{p}_{x}^{2}-\hat{p}_{y}^{2})
G^{0(R)}(-{\bf p};-(\epsilon_{n}+\omega))
G^{0(R)}({\bf p}; \epsilon_{n})\nonumber \\
&&\times \Delta_{s}^{*}({\bf R},\omega)
\nonumber \\
&=&0\;,
\end{eqnarray}

\begin{eqnarray}
\Delta_{Ic,4}^{*,N(R)}({\bf R}, {\bf k};\omega)
&=&T\sum_{\epsilon_{n}\geq 0}\int \frac{d{\bf p}}{(2\pi)^{2}} 
V_{d}(\hat{k}_{x}^{2}-\hat{k}_{y}^{2})(\hat{p}_{x}^{2}-\hat{p}_{y}^{2})
G^{0(R)}(-{\bf p};-(\epsilon_{n}+\omega))
G^{0(R)}({\bf p}; \epsilon_{n})\nonumber \\
&& \times 
\Delta_{d}^{*}({\bf R},\omega)(\hat{p}_{x}^{2}-\hat{p}_{y}^{2})
\nonumber \\
&=&
\frac{V_{d}N(0)}{4}
\{ [\ln\frac{2e^{\gamma}\omega_{D}}{\pi T}+\psi(\frac{1}{2})-
\psi(\frac{1}{2}+\frac{\rho_{1}}{2})] 
+\frac{i\omega}{4\pi T}\psi^{\prime}(\frac{1}{2}+\frac{\rho_{1}}{2}
)\}\nonumber \\
&&\times
\Delta_{d}^{*}({\bf R},\omega)
(\hat{k}_{x}^{2}-\hat{k}_{y}^{2})
\;,
\end{eqnarray}
\end{mathletters}
where
$\gamma$ is the Euler constant,
$\omega_{0}$ is the cut-off frequency, $\rho_{s}=1/\pi T\tau_{s}$,
$\rho_{1}=1/2\pi \tau_{1}$, $\psi^{\prime} (x)$ is the derivative of 
the digamma function $\psi(x)$.

As for the results of $\Delta_{Ig}^{*,N(R)}=
\sum_{i=1}^{4} \Delta_{Ig,i}^{*,N(R)}$  and 
$\Delta_{II}^{*,N(R)}=\sum_{i=1}^{16} \Delta_{II,i}^{*,N(R)}$,
the second term in Eq.~(\ref{EQ:renorm1}) can be dropped since it gives a 
very small higher order correction due to the fact that 
the gap function usually has a temporal variation over a time scale very 
long compared to the range of the Green function, that is,   
$\omega \tau_{1}\ll 1$. Therefore, the details to evaluate them are 
the same as the static case.~\cite{XRT96,KT97} Here we just 
give the results 
\begin{eqnarray}
\Delta^{*,N(R)}_{I1g}({\bf R},{\bf k};\omega)&=
&-\frac{V_{s}N(0)}{8}(\frac{v_{F}}{\pi T})^{2}[\chi_{2,1}{\bf
\Pi}^{2}\Delta^{*}_{s}({\bf
R};\omega^{\prime})+\frac{1}{2}\chi_{1,2}(\Pi_{x}^{2}-\Pi_{y}^{2})
\Delta_{d}^{*}({\bf R};\omega^{\prime})]\nonumber \\
&&
-\frac{V_{d}(\hat{k}_{x}^{2}-\hat{k}_{y}^{2})N(0)}{16}(\frac{v_{F}}{\pi
T})^{2}[\chi_{1,2}(\Pi_{x}^{2}-\Pi_{y}^{2})
\Delta_{s}^{*}({\bf R};\omega^{\prime})
\nonumber \\
&&+\chi_{0,3}{\bf \Pi}^{2}\Delta^{*}_{d}({\bf R};\omega^{\prime})]
\label{EQ:OD-G}
\end{eqnarray}
\begin{eqnarray}
\Delta_{II}^{*,N(R)}({\bf R},{\bf k};\omega)&=&-\frac{V_{s}N(0)}{2(\pi 
T)^{2}} \{ [\chi_{3,0}-\rho_{s}\chi_{4,0}]
\Delta^{*}_{s}({\bf R};\omega^{\prime})\Delta_{s}({\bf
R};\omega^{\prime\prime})\Delta^{*}_{s}({\bf
R};\omega^{\prime\prime\prime})\nonumber \\
&&+[\chi_{2,1}-\frac{\rho_{1}}{2}
\chi_{2,2}] \Delta^{*}_{d}({\bf R};\omega^{\prime})\Delta_{d}({\bf
R};\omega^{\prime\prime})\Delta^{*}_{s}({\bf
R};\omega^{\prime\prime\prime})
\nonumber \\
&&+\frac{1}{2}\chi_{2,1}\Delta^{*}_{d}({\bf
R};\omega^{\prime})\Delta_{s}({\bf
R};\omega^{\prime\prime})\Delta^{*}_{d}({\bf
R};\omega^{\prime\prime\prime}]
\nonumber \\
&&-\frac{V_{d}(\hat{k}_{x}^{2}-\hat{k}_{y}^{2})N(0)}{4(\pi T)^{2}}
\{ \chi_{2,1}
\Delta^{*}_{s}({\bf R};\omega^{\prime})\Delta_{d}({\bf
R};\omega^{\prime\prime})\Delta^{*}_{s}({\bf
R};\omega^{\prime\prime\prime})
\nonumber \\
&&+[\chi_{2,1}-\frac{\rho_{1}}{2}\chi_{2,2}]
\Delta^{*}_{d}({\bf R};\omega^{\prime})\Delta_{s}({\bf
R};\omega^{\prime\prime})\Delta^{*}_{s}({\bf
R};\omega^{\prime\prime\prime})
\nonumber \\
&&+\frac{3}{4}\chi_{0,3}\Delta^{*}_{d}({\bf
R};\omega^{\prime})\Delta_{d}({\bf
R};\omega^{\prime\prime})\Delta^{*}_{d}({\bf
R};\omega^{\prime\prime\prime}]\}\;,
\label{EQ:OD-II}
\end{eqnarray}
where 
$v_{F}$
is the Fermi velocity, and $\chi_{m,m^{\prime}}$ is a function defined as
\begin{equation}
\chi_{m,m^{\prime}}=\sum_{n\ge 0} \frac{1}{(2n+1+\rho_{s})^{m}
(2n+1+\rho_{1})^{m^{\prime}}}\;.
\end{equation}

%In the evaluation of $\Delta^{*}_{I2}$, we
%should introduce an additional renormalization factor for
%$\langle G^{0(R)}(\epsilon)G^{0(A)}(\epsilon-\omega)\rangle$ and
%$\langle G^{0(R)}(-\epsilon)G^{0(A)}(-(\epsilon-\omega))\rangle$,
%\begin{equation}
%I_{D}=\frac{i}{\omega\tau_{1}}\;.
%\end{equation}
%On the other hand, when $T_{c}-T\ll T_{c}$ and $\omega\ll T_{c}$,
%$\tanh(\epsilon/2T)-\tanh[(\epsilon-\omega^{\prime})/2T]\approx
%(\omega^{\prime}/2T)\cosh^{-2}(\epsilon/2T)$.

\subsection{Anomalous part}
The anomalous part contains integrals of the products of the retarded and 
advanced Green functions, and is therefore sensitive to the details of 
the spectrum. Following Ref.~\cite{Gor68},
we summarize here the diagrammatic rule for the evaluation of this part. 
In each diagram, the solid (electron) lines forming the upper part of the 
diagram correspond to the retarded Green function $G^{0(R)}({\bf p};\epsilon)=
[\epsilon-\xi_{\bf p}+i/2\tau_{1}]^{-1}$ for those lines with arrows to 
the right and to $G^{0(R)}(-{\bf p};-\epsilon)=
[-\epsilon-\xi_{-{\bf p}}-i/2\tau_{1}]^{-1}$ for those with arrows to the 
left. The solid lines in the lower part of the diagram correspond to the 
advanced Green function 
$G^{0(A)}({\bf p};\epsilon)=
[\epsilon-\xi_{\bf p}-i/2\tau_{1}]^{-1}$
for those lines with arrows to the left and to 
$G^{0(A)}(-{\bf p};-\epsilon)=
[-\epsilon-\xi_{-{\bf p}}+i/2\tau_{1}]^{-1}$
for those with arrows to the right. 
The triangle and the thin wavy 
line represent the order parameter $\Delta$ 
and the vertex interaction with the electromagnetic field, respectively. 
The dashed line corresponds to the impurity scattering.    
If the dashed line encompasses an even number of $\Delta$, a 
factor $1/2\pi \tau_{1}N(0)$ should be assigned. If it encompasses an 
odd number of $\Delta$, a factor $1/2\pi \tau_{2}N(0)$ should be assigned.  

As shown in Fig.~\ref{FIG:I}, the staircase which is the summation of 
the ladder diagrams, has a singular value.  
We denote it by $I(\omega,{\bf k})$, which satisfies a ladder-type equation
\begin{eqnarray}
I(\omega,{\bf k})&=&\frac{1}{2\pi \tau_{1} N(0)}\{ 1+\int d{\bf p} 
G^{0(R)}({\bf p};\epsilon)G^{0(A)}({\bf p}-{\bf k}; \epsilon-\omega)
I(\omega,{\bf k})\}\nonumber \\
&=&\frac{1}{2\pi \tau_{1} N(0)}\{ 1+N(0)\int \int  \frac{d\xi d\theta/2\pi}
{(\epsilon-\xi+\frac{i}{2\tau_{1}})(\epsilon-\omega+{\bf 
v}_{F}\cdot {\bf k}-\frac{i}{2\tau_{1}} )} 
I(\omega,{\bf k})\} \nonumber \\
&=&\frac{1}{2\pi \tau_{1} N(0)}\{ 1+N(0)\int  \frac{(2\pi i) d\theta/2\pi}
{\omega-{\bf v}_{F}\cdot {\bf k}+\frac{i}{\tau_{1}}  } 
I(\omega,{\bf k})\} \;.
\end{eqnarray}
Under the condition $\omega \tau_{1}\ll 1$ and $v_{F}k\tau_{1}\ll 1$,
we obtain 
\begin{equation}
\label{EQ:I}
I(\omega,{\bf k})=\frac{1}{2\pi 
\tau_{1}N(0)}\frac{1}{(-i\omega+Dk^{2})\tau_{1}}\;,
\end{equation}
where $D=v_{F}^{2}\tau_{1}/2$ is the difussion constant for the two 
dimensional systems. In the real time and coordinate space, 
$I^{-1}$ is proportional to the operator 
$ \frac{\partial}{\partial t}-D\nabla^{2}$.    
It is important to note that the denominator of Eq.~(\ref{EQ:I}) can be 
very small if $\omega \tau_{1}$ and $v_{F}k\tau_{1}$ are both small. 
This fact 
makes it necessary to sum additionally diagrams containing arbitrary 
number of staircases $I(\omega,{\bf k})$, separated by parts including 
$\Delta$ and $\Delta^{*}$. Under the assumption $\tau_{s}\Delta_{s}\ll 1 $
and $\tau_{1}\Delta_{1}\ll 1 $, we need only be concerned with those
diagrams of the order of $\Delta^{2}$. These diagrams together lead to the 
diffusion equation for the vertex parts $\Gamma^{+}$ and $\Gamma^{-}$ 
as shown in Fig.~\ref{FIG:GAMMA}.     

The kernel $Q_{1}$ corresponds to the diagrams shown in 
Fig.~\ref{FIG:Q1} and is given by 
\begin{eqnarray}
Q_{1}({\bf R};\omega)=Q_{1}^{(a)}({\bf R};\omega)
+Q_{1}^{(b)}({\bf R};\omega)+Q_{1}^{(c)}({\bf R};\omega)
\end{eqnarray}
with 
\begin{eqnarray}
Q_{1}^{(a)}({\bf R};\omega)&=&-\int \frac{d{\bf p}}{(2\pi)^{2} }
[G^{0(R)}({\bf p};\epsilon)]^{2} 
G^{0(R)}(-{\bf p};-\epsilon)
G^{0(A)}({\bf p};\epsilon)\nonumber \\
&&\times \tilde{\Delta}_{\omega^{\prime}}({\bf R},{\bf p})
\tilde{\Delta}_{\omega}^{*}({\bf R},{\bf p})\;,
\end{eqnarray}
\begin{eqnarray}
Q_{1}^{(b)}({\bf R};\omega)&=&-\frac{1}{2\pi\tau_{2}N(0)} 
\int \frac{d{\bf p}}{(2\pi)^{2}} G^{0(R)}({\bf p};\epsilon) 
G^{0(R)}(-{\bf p};-\epsilon)G^{0(A)}({\bf p};\epsilon)
\tilde{\Delta}_{\omega^{\prime}}({\bf R},{\bf p})
\nonumber \\
&&\times \int \frac{d{\bf p}^{\prime}}{(2\pi)^{2}} 
G^{0(R)}({\bf p}^{\prime};\epsilon) G^{0(R)}(-{\bf p}^{\prime};-\epsilon)
G^{0(A)}({\bf p}^{\prime};\epsilon)
\tilde{\Delta}_{\omega}^{*}({\bf R},{\bf p}^{\prime})\;,
\end{eqnarray}
\begin{eqnarray}
Q_{1}^{(c)}({\bf R};\omega)&=&-\frac{1}{2\pi\tau_{1}N(0)} 
\int \frac{d{\bf p}^{\prime}}{(2\pi)^{2}}[G^{0(R)}({\bf p}^{\prime};\epsilon
)]^{2} G^{0(A)}({\bf p}^{\prime};\epsilon) 
\nonumber \\
&&\times \int \frac{d{\bf p}}{(2\pi)^{2}} [G^{0(R)}({\bf p};\epsilon)]^{2} 
G^{0(R)}(-{\bf p};-\epsilon)
\tilde{\Delta}_{\omega^{\prime}}({\bf R},{\bf p})
\tilde{\Delta}_{\omega}^{*}({\bf R},{\bf p})\;.
\end{eqnarray}
Here for simplicity of notation, the vortex renormalization factor is 
included in the order parameter. For the $d$-wave component, it is 
unrenormalized, i.e., $\tilde{\Delta}_{d}=\Delta_{d}$; while for the 
$s$-wave component, $\tilde{\Delta}_{s}=\eta^{(R,A)} \Delta_{s}$ or 
$\Delta_{s}$  depends on whether the vertex connects only retarded and 
only advanced Green functions or it connects a retarded and an advanced 
Green function.     

In view of the specific form of the $s$-wave and $d$-wave pairing states, 
we see that the contribution from the $s$-wave and $d$-wave component is 
decoupled. Therefore, we obtain 
\begin{eqnarray}
Q_{1}({\bf R};\omega)&=&Q_{1,s}({\bf R};\omega)+Q_{1,d}({\bf R};\omega)
\nonumber \\
&=&-\frac{\pi i\tau_{1}^{2}N(0)}{\epsilon +\frac{i}{\tau_{s}}} 
\Delta_{s}({\bf R};\omega^{\prime})\Delta_{s}^{*}({\bf R};\omega)
-\frac{\pi i\tau_{1}^{2}N(0)}{2(\epsilon +\frac{i}{2\tau_{1}})} 
\Delta_{d}({\bf R};\omega^{\prime})\Delta_{d}^{*}({\bf R};\omega)\;.
\label{EQ:Q1}
\end{eqnarray}

We can find $Q_{2}$ from $Q_{1}$ by merely replacing all explicit $i$'s by 
$-i$'s 
\begin{equation}
\label{EQ:Q2}
Q_{2}({\bf R};\omega)=\frac{\pi i\tau_{1}^{2}N(0)}{\epsilon -\frac{i}{\tau_{s}}} 
\Delta_{s}({\bf R};\omega^{\prime})\Delta_{s}^{*}({\bf R};\omega)
+\frac{\pi i\tau_{1}^{2}N(0)}{2(\epsilon -\frac{i}{2\tau_{1}})} 
\Delta_{d}({\bf R};\omega^{\prime})\Delta_{d}^{*}({\bf R};\omega)
\end{equation}

The diagram shown in Fig.~\ref{FIG:Q3} leads to 
\begin{equation}
Q_{3}({\bf R};\omega)=Q_{3}^{(a)}({\bf R};\omega)+Q_{3}^{(b)}({\bf R};\omega)
+Q_{3}^{(c)}({\bf R};\omega)\;,
\end{equation}
with 
\begin{eqnarray}
Q_{3}^{(a)}({\bf R};\omega)&=&\int \frac{d{\bf p}}{(2\pi)^{2} }
G^{0(R)}({\bf p};\epsilon) 
G^{0(R)}(-{\bf p};-\epsilon)
G^{0(A)}({\bf p};\epsilon)G^{0(A)}(-{\bf p};-\epsilon)
\nonumber \\
&&\times \tilde{\Delta}_{\omega^{\prime}}({\bf R},{\bf p})
\tilde{\Delta}_{\omega}^{*}({\bf R},{\bf p})\;,
\end{eqnarray}
\begin{eqnarray}
Q_{3}^{(b)}({\bf R};\omega)&=&\frac{1}{2\pi\tau_{2}N(0)} 
\int \frac{d{\bf p}}{(2\pi)^{2}} G^{0(R)}({\bf p};\epsilon) 
G^{0(R)}(-{\bf p};-\epsilon)G^{0(A)}(-{\bf p};-\epsilon)
\tilde{\Delta}_{\omega^{\prime}}({\bf R},{\bf p})
\nonumber \\
&&\times \int \frac{d{\bf p}^{\prime}}{(2\pi)^{2}} G^{0(R)}({\bf p}^{\prime};
\epsilon ) 
G^{0(A)}({\bf p}^{\prime};\epsilon)
G^{0(A)}(-{\bf p}^{\prime};-\epsilon)
\tilde{\Delta}_{\omega}^{*}({\bf R},{\bf p}^{\prime})\;,
\end{eqnarray}
\begin{eqnarray}
Q_{3}^{(c)}({\bf R};\omega)&=&\frac{1}{2\pi\tau_{2}N(0)} 
\int \frac{d{\bf p}}{(2\pi)^{2}} G^{0(R)}({\bf p};\epsilon) 
G^{0(R)}(-{\bf p};-\epsilon)G^{0(A)}({\bf p};\epsilon)
\tilde{\Delta}_{\omega^{\prime}}({\bf R},{\bf p})
\nonumber \\
&&\times \int \frac{d{\bf p}^{\prime}}{(2\pi)^{2}} 
G^{0(R)}(-{\bf p}^{\prime};-\epsilon) 
G^{0(A)}({\bf p}^{\prime};\epsilon)
G^{0(A)}(-{\bf p}^{\prime};-\epsilon)
\tilde{\Delta}_{\omega}^{*}({\bf R},{\bf p}^{\prime})\;.
\end{eqnarray}
The  algebra gives 
\begin{equation}
\label{EQ:Q3}
Q_{3}({\bf R};\omega)=\frac{2\pi N(0)\tau_{1}^{2}/\tau_{s}}{\epsilon^{2}
+\tau_{s}^{-2}}
\Delta_{s}({\bf R};\omega_{1})\Delta_{s}^{*}({\bf R};\omega_{2})
+\frac{2\pi N(0)\tau_{1}^{2}}{2[\epsilon^{2}+(2\tau_{1})^{-2}]}
\Delta_{d}({\bf R};\omega_{1})\Delta_{d}^{*}({\bf R};\omega_{2})\;.
\end{equation}
From the results of $Q_{1,2,3}$ given by Eqs.~(\ref{EQ:Q1}), 
(\ref{EQ:Q2}), and (\ref{EQ:Q3}), it is not difficult to prove the
relation 
\begin{equation}
Q_{3}({\bf R};\omega)=-(Q_{1}({\bf R};\omega)+Q_{2}({\bf R};\omega))\;.
\end{equation}

In addition, the other three separate terms are as follows: 
\begin{eqnarray}
S_{1}({\bf R};\omega)&=&-\int \frac{d{\bf p}}{(2\pi)^{2} }
G^{0(R)}({\bf p};\epsilon) 
G^{0(R)}(-{\bf p};-\epsilon)
G^{0(A)}({\bf p};\epsilon)\nonumber \\
&&\times (\tanh 
\frac{\epsilon}{2T}-\tanh \frac{\epsilon-\omega_{2}}{2T})
\tilde{\Delta}_{\omega_{1}}({\bf R},{\bf p})
\Delta_{\omega_{2}}^{*}({\bf R},{\bf p})\;,\nonumber \\
&=&\frac{1}{2T}\cosh^{-2}\frac{\epsilon}{2T}
[\frac{\pi\tau_{1}N(0)}{\epsilon+i/\tau_{s}}\Delta_{s}({\bf 
R};\omega_{1})\omega_{2}\Delta_{s}^{*}({\bf R};\omega_{2})\nonumber \\
&&
+\frac{\pi\tau_{1}N(0)}{2(\epsilon+i/2\tau_{1})}\Delta_{d}({\bf 
R};\omega_{1})\omega_{2}\Delta_{d}^{*}({\bf R};\omega_{2})\;,
\end{eqnarray}
\begin{eqnarray}
S_{2}({\bf R};\omega)&=&-\int \frac{d{\bf p}}{(2\pi)^{2} }
G^{0(R)}({\bf p};\epsilon) 
G^{0(A)}(-{\bf p};-\epsilon)
G^{0(A)}({\bf p};\epsilon)\nonumber \\
&&\times 
(\tanh \frac{\epsilon}{2T}-\tanh \frac{\epsilon-\omega_{1}}{2T})
\Delta_{\omega_{1}}({\bf R},{\bf p})
\tilde{\Delta}_{\omega_{2}}^{*}({\bf R},{\bf p})\;,\nonumber \\
&=&\frac{1}{2T}\cosh^{-2}\frac{\epsilon}{2T}
[\frac{\pi\tau_{1}N(0)}{\epsilon-i/\tau_{s}}\Delta_{s}({\bf 
R};\omega_{1})\omega_{2}\Delta_{s}^{*}({\bf R};\omega_{2})
\nonumber \\
&&
+\frac{\pi\tau_{1}N(0)}{2(\epsilon-i/2\tau_{1})}\Delta_{d}({\bf 
R};\omega_{1})\omega_{2}\Delta_{d}^{*}({\bf R};\omega_{2})\;,
\end{eqnarray}
and 
\begin{eqnarray}
S_{3}({\bf R};\omega)&=&\int \frac{d{\bf p}}{(2\pi)^{2}} 
(\tanh \frac{\epsilon}{2T}-\tanh \frac{\epsilon-\omega}{2T})
G^{0(R)}({\bf p};\epsilon)G^{0(A)}({\bf p};\epsilon)
e\varphi_{\omega}({\bf R})\nonumber \\
&=&\frac{\pi 
i\tau_{1}N(0)}{2T}\cosh^{-2}\frac{\epsilon}{2T}(-i\omega)
e\varphi_{\omega}({\bf R}) \;.
\end{eqnarray}
Here we have approximated 
\begin{equation}
\tanh \frac{\epsilon}{2T}-\tanh \frac{\epsilon-\omega}{2T} \approx 
\frac{\omega}{2T} \cosh^{-2} \frac{\epsilon}{2T}\;,
\end{equation}
when  $\omega \ll T$.

From the results for $Q$'s and $S$'s, we obtain the diffusion equation 
for $\Gamma^{+}$
\begin{mathletters}
\begin{eqnarray}
(\frac{\partial }{\partial t}-D\nabla^{2})\Gamma^{+}&=& 
-\frac{1}{2\pi \tau_{1}^{2}N(0)}[(\sum_{i=1}^{3}S_{i}
+\sum_{i=1}^{2}\Gamma^{+})-Q_{3}\Gamma^{-}]\nonumber \\
&=& \frac{1}{4T\tau_{1}}\cosh^{-2}\frac{\epsilon}{2T}\{- 
\frac{i\epsilon}{\epsilon^{2}+\tau_{s}^{-2}}\frac{\partial \vert 
\Delta_{s}\vert^{2}}{\partial t}
- \frac{i\epsilon}{2[\epsilon^{2}+(2\tau_{1})^{-2}]}\frac{\partial \vert 
\Delta_{d}\vert^{2}}{\partial t}
-2ie\frac{\partial \varphi}{\partial t}\} 
\nonumber \\
&&-\frac{\tau_{s}^{-1}}{\epsilon^{2}+\tau_{s}^{-2}}
(\Delta_{s}\frac{\partial 
\Delta_{s}^{*}}{\partial t}-\Delta_{s}^{*}\frac{\partial 
\Delta_{s}}{\partial t})
-\frac{(2\tau_{1})^{-1}}{2[\epsilon^{2}+(2\tau_{1})^{-2}]}
(\Delta_{d}\frac{\partial \Delta_{d}^{*}}{\partial t}
-\Delta_{d}^{*}\frac{\partial 
\Delta_{d}}{\partial t})
\nonumber \\
&&-\{ \frac{\tau_{s}^{-1}}{\epsilon^{2}+\tau_{s}^{-2}}\vert 
\Delta_{s}\vert^{2}
+\frac{(2\tau_{1})^{-1}}{2[\epsilon^{2}+(2\tau_{1})^{-2}]}\vert\Delta_{d}
\vert^{2}\} (\Gamma^{+}+\Gamma^{-})\;,
\end{eqnarray}
where  $\Delta_{s,d}$ and $\varphi$ are functions of 
${\bf R}$ and $t$. and $\Gamma^{\pm}$ are funtions of ${\bf R}$, $t$, and 
$\epsilon$. 

Similary,  the diffusion equation for $\Gamma^{-}$ is found to be 
\begin{eqnarray}
(\frac{\partial }{\partial t}-D\nabla^{2})\Gamma^{-}&=& 
\frac{1}{4T\tau_{1}}\cosh^{-2}\frac{\epsilon}{2T}\{ 
\frac{i\epsilon}{\epsilon^{2}+\tau_{s}^{-2}}\frac{\partial \vert 
\Delta_{s}\vert^{2}}{\partial t}+
\frac{i\epsilon}{2[\epsilon^{2}+(2\tau_{1})^{-2}]}\frac{\partial \vert 
\Delta_{d}\vert^{2}}{\partial t}
-2ie\frac{\partial \varphi}{\partial t}\} 
\nonumber \\
&&-\frac{\tau_{s}^{-1}}{\epsilon^{2}+\tau_{s}^{-2}}
(\Delta_{s}\frac{\partial 
\Delta_{s}^{*}}{\partial t}-\Delta_{s}^{*}\frac{\partial 
\Delta_{s}}{\partial t})
-\frac{(2\tau_{1})^{-1}}{2[\epsilon^{2}+(2\tau_{1})^{-2}]}
(\Delta_{d}\frac{\partial \Delta_{d}^{*}}{\partial t}
-\Delta_{d}^{*}\frac{\partial 
\Delta_{d}}{\partial t})
\nonumber \\
&&-\{\frac{\tau_{s}^{-1}}{\epsilon^{2}+\tau_{s}^{-2}}\vert 
\Delta_{s}\vert^{2}
+\frac{(2\tau_{1})^{-1}}{2[\epsilon^{2}+(2\tau_{1})^{-2}]}\vert\Delta_{d}
\vert^{2}\}(\Gamma^{+}+\Gamma^{-})\;.
\end{eqnarray}
\end{mathletters}
  
These two diffusion equations can be rewritten as
\begin{mathletters}
\label{EQ:TDGL-DF}
\begin{equation}
(\frac{\partial}{\partial t}-D\nabla^{2})(\Gamma^{+}-\Gamma^{-})=
-\frac{i\cosh^{-2}(\frac{\epsilon}{2T})}{2T\tau_{1}}
\{\frac{\epsilon}{\epsilon^{2}+\tau_{s}^{-2}}\frac{\partial 
\vert\Delta_{s}\vert^{2}}{\partial t}+
\frac{\epsilon}{2[\epsilon^{2}+(2\tau_{1})^{-2}]}
\frac{\partial \vert\Delta_{d}\vert^{2}}{\partial t}\}\;, \label{EQ:TDGL5}
\end{equation}  
\begin{eqnarray}
(\frac{\partial}{\partial t}-D\nabla^{2})(\Gamma^{+}+\Gamma^{-})&=&
-\frac{\cosh^{-2}(\frac{\epsilon}{2T})}{2T\tau_{1}}
\{\frac{\tau_{s}^{-1}}{\epsilon^{2}+\tau_{s}^{-2}}
(\Delta_{s}\frac{\partial\Delta_{s}^{*}}{\partial t}
-\Delta_{s}^{*}\frac{\partial\Delta_{s}}{\partial t}) \nonumber \\
&&+\frac{(2\tau_{1})^{-1}}{2[\epsilon^{2}+(2\tau_{1})^{-2}]}
( \Delta_{d}\frac{\partial\Delta_{d}^{*}}{\partial t}
-\Delta_{d}^{*}\frac{\partial\Delta_{d}}{\partial t})
-2ie\frac{\partial\varphi}{\partial t}
\}\nonumber \\
&&-\{\frac{2\tau_{s}^{-1}}{\epsilon^{2}+\tau_{s}^{-2}}
\vert \Delta_{s} \vert^{2}+
\frac{(2\tau_{1})^{-1}}{\epsilon^{2}+(2\tau_{1})^{-2}}
\vert \Delta_{d} \vert^{2} \} (\Gamma^{+}+\Gamma^{-})
\;. \label{EQ:TDGL6} 
\end{eqnarray} 
\end{mathletters}

With the results of $\Gamma^{\pm}$, the anomalous part represented 
in Fig.~\ref{FIG:ANOM-OD} is given by 
\begin{eqnarray}
\Theta&=&\frac{1}{4\pi i}\{-\int d\epsilon 
\int \frac{d{\bf p}}{(2\pi)^{2}} V({\bf k}-{\bf p}) 
G^{0(R)}(-{\bf p};-\epsilon)
G^{0(R)}({\bf p};\epsilon)
G^{0(A)}({\bf p};\epsilon)
\tilde{\Delta}_{\omega_{1}}^{*}({\bf R},{\bf p})
\Gamma_{\omega_{2}}^{+}({\bf R},\epsilon) 
\nonumber \\
&&-\int d\epsilon 
\int \frac{d{\bf p}}{(2\pi)^{2}} V({\bf k}-{\bf p}) 
G^{0(R)}(-{\bf p};-\epsilon)
G^{0(A)}(-{\bf p};-\epsilon)
G^{0(A)}({\bf p};\epsilon)
\tilde{\Delta}_{\omega_{1}}^{*}({\bf R},{\bf p})
\Gamma_{\omega_{2}}^{-}({\bf R},\epsilon)\}
\nonumber \\
&=&-\frac{\tau_{1}N(0)}{4}V_{s}\Delta_{s}^{*}({\bf R};\omega_{1}) 
\int d\epsilon [
\frac{\epsilon-i\tau_{s}^{-1}}{\epsilon+\tau_{s}^{-2}}
\Gamma_{\omega_{2}}^{+}({\bf 
R},\epsilon)- 
\frac{\epsilon+i\tau_{s}^{-1}}{\epsilon+\tau_{s}^{-2}}
\Gamma_{\omega_{2}}^{-}({\bf R},\epsilon)] 
\nonumber \\ 
&&-\frac{\tau_{1}N(0)}{8}V_{d}(\hat{k}_{x}^{2}-\hat{k}_{y}^{2}) 
\Delta_{d}^{*}({\bf R};\omega_{1}) \int d\epsilon 
[\frac{\epsilon-i(2\tau_{1})^{-1}}{\epsilon+(2\tau_{1})^{-2}}
\Gamma_{\omega_{2}}^{+}({\bf R},\epsilon)
\nonumber \\
&&-\frac{\epsilon+i(2\tau_{1})^{-1}}{\epsilon+(2\tau_{1})^{-2}}
\Gamma_{\omega_{2}}^{-}({\bf R},\epsilon)]\;.
\label{EQ:ANOMAL}
\end{eqnarray}
where $\Theta$ is a function of ${\bf R}$, $\omega$, and ${\bf k}$.

\subsection{TDGL equations for the order parameters}
From Eqs.~(\ref{EQ:OD-C}), (\ref{EQ:OD-G}), and (\ref{EQ:OD-II}), and 
(\ref{EQ:ANOMAL}), by performing the inverse Fourier transform 
and comparing both sides of 
the gap function for $\hat{\bf k}$-independent terms and the terms 
proportional to $\hat{k}_{x}^{2}-\hat{k}_{y}^{2}$, we 
obtain the coupled TDGL equations for the order parameter components: 
\begin{eqnarray}
&-\frac{1}{\gamma_{s}}
\frac{\partial \Delta_{s}^{*}({\bf R},t)}{\partial t}
+2\Phi_{s}({\bf R},t)  
\Delta_{s}^{*}({\bf R},t)= &
\nonumber \\
&
2\alpha_{s}\Delta_{s}^{*}({\bf R},t)
+(\frac{v_{F}}{\pi T})^{2}[\frac{1}{2}\chi_{2,1}{\bf
\Pi}^{2}\Delta^{*}_{s}({\bf
R},t)+\frac{1}{4}\chi_{1,2}(\Pi_{x}^{2}-\Pi_{y}^{2})
\Delta_{d}^{*}({\bf R},t)]&\nonumber \\
&+(\frac{1}{\pi T})^{2}\{
2(\chi_{3,0}-\rho_{s}\chi_{4,0})
\Delta_{s}^{*}({\bf R},t)\vert\Delta_{s}({\bf R},t)\vert^{2}&
\nonumber \\
&+2(\chi_{2,1}-\frac{\rho_{1}}{2}\chi_{2,2})\vert \Delta_{d}({\bf
R},t)\vert^{2}\Delta_{s}^{*}({\bf R},t)+\chi_{2,1}
\Delta_{d}^{*2}({\bf R},t)\Delta_{s}({\bf R},t)\}\;,&\label{EQ:TDGL-S}
\end{eqnarray}
\begin{eqnarray}
&-\frac{1}{\gamma_{d}} 
\frac{\partial \Delta_{d}^{*}({\bf R},t)}{\partial t}
+2\Phi_{d}({\bf R},t)
\Delta_{d}^{*}({\bf R},t) =&
\nonumber   \\
&\alpha_{d}\Delta_{d}^{*}({\bf R},t)
+(\frac{v_{F}}{2\pi T})^{2}[\chi_{0,3}{\bf
\Pi}^{2}\Delta^{*}_{d}({\bf R},t)+\chi_{1,2}(\Pi_{x}^{2}-\Pi_{y}^{2})
\Delta_{s}^{*}({\bf R},t)]&\nonumber \\
&+(\frac{1}{\pi T})^{2}\{
(\frac{3}{4}\chi_{0,3}\Delta_{d}^{*}({\bf R},t)\vert\Delta_{d}({\bf
R},t)\vert^{2}& \nonumber \\
&+2(\chi_{2,1}-\frac{\rho_{1}}{2}\chi_{2,2})
\vert \Delta_{s}({\bf R},t)\vert^{2}\Delta_{d}^{*}({\bf R},t)+\chi_{2,1}
\Delta_{s}^{*2}({\bf R},t)\Delta_{d}({\bf R},t)\}\;.&\label{EQ:TDGL-D}
\end{eqnarray}
Here $\gamma_{s,d}$ are  two relaxation rates defined by
\begin{equation}
\gamma_{s}^{-1}=\frac{1}{2\pi T} \psi^{\prime}(\frac{1}{2}
+\frac{\rho_{s}}{2})\;,
\end{equation}
and 
\begin{equation}
\gamma_{d}^{-1}=\frac{1}{4\pi T}\psi^{\prime}(\frac{1}{2}+
\frac{\rho_{1}}{2})\;.
\end{equation}
Two quantities $\Phi_{s,d}$ are   
given by 
\begin{equation}
\mu_{s}({\bf R},t)=-\frac{i\tau_{1}}{4} \int d\epsilon [
\frac{\epsilon-i\tau_{s}^{-1}}{\epsilon^{2}+\tau_{s}^{-2}}\Gamma^{+}({\bf 
R},t,\epsilon)
-\frac{\epsilon+i\tau_{s}^{-1}}{\epsilon^{2}+\tau_{s}^{-2}}
\Gamma^{-}({\bf R},t,\epsilon)]
\;, 
\end{equation}
\begin{equation}
\mu_{d}({\bf R},t)=-\frac{i\tau_{1}}{8} \int d\epsilon 
[\frac{\epsilon-i(2\tau_{1})^{-1}}{\epsilon^{2}+(2\tau_{1})^{-2}}
\Gamma^{+}({\bf R},t,\epsilon)
-\frac{\epsilon+i(2\tau_{1})^{-1} }{\epsilon^{2}+(2\tau_{1})^{-2}}
\Gamma^{-}({\bf R},t,\epsilon)]\;. 
\end{equation}
Finally, the parameters $\alpha_{s}$ and $\alpha_{d}$ are given by  
\begin{equation}
\alpha_{s}=-[\ln \frac{T_{cs0}}{T}+\psi(\frac{1}{2})-\psi(\frac{1}{2}
+\frac{\rho_{s}}{2})]\;,
\end{equation}
and
\begin{equation}
\alpha_{d}=-[\ln \frac{T_{cd0}}{T}+\psi(\frac{1}{2})-\psi(\frac{1}{2}
+\frac{\rho_{1}}{2})]\;.
\end{equation}
$T_{cs0}$ and $T_{cd0}$ are the critical
temperatures of a clean superconductor, which are determined  by
\begin{equation}
N(0)V_{s}\ln(2e^{\gamma}\omega_{D}/\pi T_{cs0})=1\;,
\end{equation} 
and
\begin{equation}
[N(0)V_{d}/2]\ln(2e^{\gamma}\omega_{D}/\pi T_{cd0})=1\;,
\end{equation} 
with $\gamma$ the Euler constant and $\omega_{D}$ the cut-off frequency.
In the presence of impurity scatterings, two transition temperatures
are determined by the conditions $\alpha_{s}(T_{cs})=0$ and
$\alpha_{d}(T_{cd})=0$.
It is very clear that the transition temperature $T_{cs}$ for $s$-wave
order parameter can only be
affected by the magnetic impurity scattering while the transition
temperature for $d$-wave order parameter is dominantly affected by the
nonmagnetic scattering.
The critical temperature of the superconductor is defined by
$T_{c}=\mbox{max}\{T_{cs},T_{cd}\}$.
We estimate that as long as the $d$-channel interaction $V_{d}$ is larger
than about three times of the $s$-channel interaction $V_{s}$,
the pure $d$-wave state is stable in the bulk systems without
perturbations. The phase diagram of such a system in the
absence of external fields and impurities has been previously
studied in Ref.~\cite{RXT96}.

By introducing a formal free energy density  
\begin{eqnarray}
f({\bf R},t)&=&2\alpha_{s}\vert \Delta_{s}({\bf R},t)\vert^{2}
+\alpha_{d}\vert \Delta_{d}({\bf R},t)\vert^{2}
+(\frac{v_{F}}{\pi T})^{2}\{\frac{1}{2}\chi_{2,1}\vert {\bf \Pi}
\Delta^{*}_{s}({\bf R},t)\vert^{2}
+\frac{1}{4}\chi_{0,3}\vert {\bf 
\Pi}\Delta^{*}_{d}({\bf 
R},t)\vert^{2}\nonumber \\
&&+\frac{1}{4}\chi_{1,2}[ \Pi_{x}^{*}\Delta_{s}({\bf R},t)
\Pi_{x}\Delta_{d}^{*}({\bf R},t)-\Pi_{y}^{*}\Delta_{s}({\bf R},t)
\Pi_{y}\Delta_{d}^{*}({\bf R},t)+\mbox{C.C.}]\}\nonumber \\
&&+(\frac{1}{\pi T})^{2}\{ 
(\chi_{3,0}-\rho_{s}\chi_{4,0})
\vert\Delta_{s}({\bf R},t)\vert^{4}
+\frac{3}{8}(\chi_{0,3}-\frac{2\rho_{1}}{3}\chi_{0,4})
\vert\Delta_{d}({\bf R},t)\vert^{4}\nonumber \\
&&+(2\chi_{2,1}-\rho_{1}\chi_{2,2})\vert \Delta_{d}({\bf 
R},t)\vert^{2}\vert \Delta_{s}({\bf R},t)\vert^{2}
+\frac{1}{2} \chi_{2,1}
[\Delta_{d}^{*2}({\bf R},t)\Delta_{s}^{2}({\bf R},t)\nonumber \\
&&+\Delta_{s}^{*2}({\bf R},t)\Delta_{d}^{2}({\bf R},t)]\}
 %+(\frac{1}{2\pi N(0)})(\nabla\times {\bf A}({\bf R},t))^{2}\;, 
\end{eqnarray} 
the TDGL equations (\ref{EQ:TDGL-S}) and (\ref{EQ:TDGL-D}) can be written 
in a compact way 
\begin{mathletters}
\label{EQ:TDGL-OP}
\begin{equation}
-\frac{1}{\gamma_{s}}
\frac{\partial \Delta_{s}^{*}({\bf R},t)}{\partial t}
+2\Phi_{s}({\bf R},t)  
\Delta_{s}^{*}({\bf R},t) 
=\frac{\delta f({\bf R},t)}{\delta \Delta_{s}}\;,
\label{EQ:TDGL1}
\end{equation}
\begin{equation}
-\frac{1}{\gamma_{d}} 
\frac{\partial \Delta_{d}^{*}({\bf R},t)}{\partial t}
+2\Phi_{d}({\bf R},t)
\Delta_{d}^{*}({\bf R},t) 
=\frac{\delta f({\bf R},t)}{\delta \Delta_{d}}\;.
\label{EQ:TDGL2}
\end{equation}
\end{mathletters}

\section{Time-Dependent Current and Charge Density} 
\subsection{Current density}
The expression for current in ``imaginary'' frequency space is given as
\begin{equation}
{\bf J}({\bf x},\tau)=-\frac{e}{mi}(\nabla_{\bf x}-\nabla_{{\bf
x}^{\prime}})\langle \delta G({\bf x}\tau;{\bf 
x}^{\prime}\tau^{0+})\rangle \vert_{{\bf x}^{\prime}\rightarrow {\bf x}}
-\frac{2e^{2}}{m}{\bf A}({\bf x},\tau)\langle \delta G({\bf 
x}\tau;{\bf x}\tau^{0+})\rangle \;, 
\end{equation}
where $\langle \delta G_{\alpha\beta}\rangle
=\langle G_{\alpha\beta}-
 G_{\alpha\beta}^{0}\rangle=\langle \delta G\rangle \delta_{\alpha\beta}$ 
with $G_{\alpha\beta}$ defined by Eq.~(\ref{EQ:green1})
and the factor $2$ arises from the spin sum. 
Using the similar technique for the gap function,
we can divide the current into the 
normal and anomalous parts, that is,  
\begin{equation}
{\bf J}_{\omega}({\bf R})={\bf J}_{\omega}^{N}({\bf R})+{\bf 
J}_{\omega}^{A}({\bf R},t)\;. 
\end{equation}
The normal part is given by
\begin{eqnarray} 
{\bf J}_{\omega}^{N}({\bf R})&=&
(\frac{eT}{mi})\sum_{\epsilon_{n}\geq 0}
\int d{\bf R}^{\prime}d{\bf r}^{\prime}
 d{\bf R}^{\prime\prime}d{\bf r}^{\prime\prime}
[e^{-i({\bf R}^{\prime}-{\bf R})\cdot {\bf \Pi}^{*}+({\bf 
r}^{\prime}-{\bf r})\cdot \nabla_{\bf r}}\Delta_{\omega_{1}}({\bf R},{\bf 
r})] \nonumber \\
&&\times 
[e^{i({\bf R}^{\prime\prime}-{\bf R})\cdot {\bf \Pi}+({\bf 
r}^{\prime\prime}-{\bf r})\cdot \nabla_{\bf r}}\Delta_{\omega_{2}}^{*}
({\bf R},{\bf r})] G^{0(R)}({\bf R}^{\prime\prime}
-{\bf r}^{\prime\prime}/2,{\bf 
R}^{\prime}-{\bf r}^{\prime}/2;-\epsilon_{n})
\nonumber \\
&& \times
\nabla_{\bf r}\{ G^{0(R)}({\bf R}+{\bf r}/2,{\bf 
R}^{\prime}+{\bf r}^{\prime}/2;\epsilon_{n})
G^{0(R)}({\bf R}^{\prime\prime}-{\bf r}^{\prime\prime}/2,{\bf 
R}^{\prime}-{\bf r}^{\prime}/2;\epsilon_{n})\}_{{\bf r}\rightarrow 0}
\nonumber \\
&&-(\frac{eT}{mi})\sum_{\epsilon_{n}\leq 0}
\int d{\bf R}^{\prime}d{\bf r}^{\prime}
 d{\bf R}^{\prime\prime}d{\bf r}^{\prime\prime}
[e^{-i({\bf R}^{\prime}-{\bf R})\cdot {\bf \Pi}^{*}+({\bf 
r}^{\prime}-{\bf r})\cdot \nabla_{\bf r}}\Delta_{\omega_{1}}({\bf R},{\bf 
r})] \nonumber \\
&&\times 
[e^{i({\bf R}^{\prime\prime}-{\bf R})\cdot {\bf \Pi}+({\bf 
r}^{\prime\prime}-{\bf r})\cdot \nabla_{\bf r}}\Delta_{\omega_{2}}^{*}
({\bf R},{\bf r})] G^{0(A)}({\bf R}^{\prime\prime}
-{\bf r}^{\prime\prime}/2,{\bf 
R}^{\prime}-{\bf r}^{\prime}/2;-\epsilon_{n})
\nonumber \\
&& \times
\nabla_{\bf r}\{ G^{0(A)}({\bf R}+{\bf r}/2,{\bf 
R}^{\prime}+{\bf r}^{\prime}/2;\epsilon_{n})
G^{0(A)}({\bf R}^{\prime\prime}-{\bf r}^{\prime\prime}/2,{\bf 
R}^{\prime}-{\bf r}^{\prime}/2;\epsilon_{n})\}_{{\bf r}\rightarrow 0}\;.
\end{eqnarray}
The computation of this part is the same as the static 
case~\cite{XRT96,KT97} and we give as a result 
\begin{eqnarray}
{\bf J}_{\omega}^{N}({\bf R})&=& 
\frac{eE_{F}N(0)}{m(\pi T)^{2}}
\{ \frac{1}{2} \chi_{2,1}\Delta^{*}_{s}({\bf R};\omega_{2}){\bf 
\Pi}^{*}\Delta_{s}({\bf 
R};\omega_{1}) +\frac{1}{4}\chi_{0,3}\Delta^{*}_{d}({\bf R};\omega_{2}){\bf 
\Pi}^{*} \Delta_{d}({\bf R};\omega_{1})\nonumber \\
&&+\frac{1}{4}\chi_{1,2}[\Delta^{*}_{s}({\bf R};\omega_{2})\Pi_{x}^{*}
\Delta_{d}({\bf R};\omega_{1})+\Delta^{*}_{d}({\bf R};\omega_{2})\Pi_{x}^{*}
\Delta_{s}({\bf R};\omega_{1})]{\bf e}_{x} \nonumber \\
&&-
\Delta^{*}_{s}({\bf R};\omega_{2})\Pi_{y}^{*}
\Delta_{d}({\bf R};\omega_{1})+\Delta^{*}_{d}({\bf R};\omega_{2})\Pi_{y}^{*}
\Delta_{s}({\bf R};\omega_{1})]{\bf e}_{y}]\}+\mbox{C.C.}\;.
\label{EQ:CUR-N}
\end{eqnarray}
Here ${\bf e}_{x,y}$ is the unit vector along the $x(y)$-direction.

The anomalous part is represented by the diagram shown 
in Fig.~\ref{FIG:ANOM-J}. The 
contribution from the first term is 
\begin{eqnarray}
{\bf J}_{\omega}^{A,1}({\bf R})&=&-(\frac{2e}{mi})\frac{1}{4\pi i}\int 
d\epsilon (\tanh \frac{\epsilon}{2T}-\tanh \frac{\epsilon-\omega}{2T}) 
\int d{\bf x}^{\prime\prime}  [-e\varphi_{\omega}({\bf 
x}^{\prime\prime})+ \frac{e}{m}{\bf A}_{\omega}({\bf x}^{\prime\prime})\cdot 
{\bf p}_{{\bf x}^{\prime\prime}}]\nonumber \\
&&\times \nabla_{\bf r} [G^{0(R)}({\bf R}+{\bf r}/2,{\bf 
x}^{\prime\prime};\epsilon)G^{0(A)}({\bf x}^{\prime\prime},{\bf R}-{\bf 
r}/2;\epsilon)]_{{\bf r}\rightarrow 0}
\nonumber \\
&\approx& 
-(\frac{2e}{mi})\frac{1}{4\pi i}\int 
d\epsilon (\tanh \frac{\epsilon}{2T}-\tanh \frac{\epsilon-\omega}{2T}) 
\int d{\bf x}^{\prime\prime} \int \frac{d{\bf p}_{1}d{\bf 
p}_{2}}{(2\pi)^{4}} G^{0(R)}({\bf p}_{1};\epsilon)
G^{0(A)}({\bf p}_{2};\epsilon)
 \nonumber \\
&&\times 
[-e\varphi_{\omega}({\bf 
x}^{\prime\prime})+ \frac{e}{m}{\bf A}_{\omega}({\bf x}^{\prime\prime})\cdot 
{\bf p}_{2}]
\nabla_{\bf r} [e^{i{\bf p}_{1}\cdot ({\bf R}+{\bf r}/2- {\bf 
x}^{\prime\prime})} e^{ i{\bf p}_{2}\cdot ({\bf x}^{\prime\prime}-{\bf 
R}+{\bf r}/2)}]_{{\bf r}\rightarrow 0}
\nonumber \\
&=&
-(\frac{2e}{mi})\frac{1}{4\pi i}\int 
d\epsilon (\tanh \frac{\epsilon}{2T}-\tanh \frac{\epsilon-\omega}{2T}) 
\int d{\bf x}^{\prime\prime} \int \frac{d{\bf p}}{(2\pi)^{2}} 
G^{0(R)}({\bf p};\epsilon)
G^{0(A)}({\bf p};\epsilon)
 \nonumber \\
&& \times {\bf p} [-e\varphi_{\omega}({\bf 
R})+ \frac{e}{m}{\bf A}_{\omega}({\bf R})\cdot {\bf p}]
\nonumber \\
&=&-\sigma [-i\omega {\bf A}_{\omega}({\bf R})]\;,
\label{EQ:CUR-A1}
\end{eqnarray}
where $\sigma=N(0)e^{2}v_{F}^{2}\tau_{1}=2N(0)e^{2}D$ is the normal-state
conductivity. Here we have used the integral 
\begin{equation}
\int d\epsilon \frac{1}{2T} \cosh^{-2}\frac{\epsilon}{2T}=2\;.
\end{equation}
Similarly, the contribution of the second term is given by 
\begin{eqnarray}
{\bf J}_{\omega}^{A,2}({\bf R})&=&-(\frac{2e}{mi})\frac{1}{4\pi i}\int 
d\epsilon  
\int d{\bf x}^{\prime\prime}  (-\Gamma_{\omega}^{+}({\bf R},\epsilon)
)\nonumber \\
&&\times \nabla_{\bf r} [G^{0(R)}({\bf R}+{\bf r}/2,{\bf 
x}^{\prime\prime};\epsilon)G^{0(A)}({\bf x}^{\prime\prime},{\bf R}-{\bf 
r}/2;\epsilon)]_{{\bf r}\rightarrow 0}
\nonumber \\
&=&-\frac{\sigma \tau_{1}}{2ie} \nabla \int d\epsilon 
\Gamma_{\omega}^{+}({\bf R},\epsilon)
\nonumber \\
&=&-\frac{\sigma \tau_{1}}{4ie} \nabla \int d\epsilon 
[\Gamma_{\omega}^{+}({\bf R},\epsilon)+
\Gamma_{\omega}^{-}({\bf R},\epsilon)]\;.
\label{EQ:CUR-A2}
\end{eqnarray}

By performing an inverse Fourier transform to Eqs.~(\ref{EQ:CUR-N}), 
(\ref{EQ:CUR-A1}), and (\ref{EQ:CUR-A2}), we obtain 
the current in real time 
\begin{equation}
\label{EQ:TDGL-J}
{\bf J}({\bf R},t)={\bf J}_{n}({\bf R},t)+{\bf J}_{s}({\bf R},t)\;.
\end{equation}
Here the normal state current is given by
\begin{equation}
{\bf J}_{n}({\bf R},t)=-\sigma[\nabla \tilde{\varphi}({\bf R},t)
+\frac{\partial {\bf A}({\bf R},t)}{\partial t}]
\end{equation}
where 
\begin{equation}
\tilde{\varphi}({\bf R},t)= \frac{\tau_{1}}{4ie}\int 
d\epsilon [\Gamma^{+}({\bf R},t,\epsilon)
+\Gamma^{-}({\bf R},t,\epsilon)]
\end{equation}
can be considered as the effective  
electro-chemical potential for quasi-particles. The supercurrent 
is given by 
\begin{eqnarray}
{\bf J}_{s}({\bf R},t)&=&\frac{eE_{F}N(0)}{m(\pi T)^{2}}
\{ \frac{1}{2} \chi_{2,1}\Delta^{*}_{s}({\bf R},t){\bf 
\Pi}^{*}\Delta_{s}({\bf 
R},t) +\frac{1}{4}\chi_{0,3}\Delta^{*}_{d}({\bf R},t){\bf \Pi}^{*}
\Delta_{d}({\bf R},t)\nonumber \\
&&+\frac{1}{4}\chi_{1,2}[\Delta^{*}_{s}({\bf R},t)\Pi_{x}^{*}
\Delta_{d}({\bf R},t)+\Delta^{*}_{d}({\bf R},t)\Pi_{x}^{*}
\Delta_{s}({\bf R},t)]{\bf e}_{x} \nonumber \\
&&-\Delta^{*}_{s}({\bf R},t)\Pi_{y}^{*}
\Delta_{d}({\bf R},t)+\Delta^{*}_{d}({\bf R},t)\Pi_{y}^{*}
\Delta_{s}({\bf R},t)]{\bf e}_{y}]\}+\mbox{C.C.} \nonumber \\
&=&-\frac{N(0)}{4}\frac{\delta f({\bf R},t)}{\delta {\bf A}}\;.
\end{eqnarray}

\subsection{Charge density}
The charge density in the ``imaginary'' time space is defined by
\begin{equation}
\rho({\bf x},\tau)=-2e \langle G({\bf x}\tau,{\bf x}\tau^{+0})\rangle \;.
\end{equation}
After the analytical continuation, we have 
\begin{eqnarray}
\rho_{\omega}({\bf x})
&=&-2eT\sum_{\epsilon}G_{\epsilon,\epsilon-\omega}({\bf x},{\bf x})
\nonumber \\ 
&=&\rho^{N}_{\omega}({\bf R})+\rho^{A}_{\omega}({\bf R})\;,
\end{eqnarray}
with 
\begin{eqnarray}
\rho^{N}_{\omega}({\bf R})&=&(-2e)\frac{1}{4\pi i}\int d\epsilon \tanh 
\frac{\epsilon}{2T} \int d{\bf x}_{1} (-e\varphi_{\omega}({\bf x}_{1})
\nonumber \\
&&\times [G^{0(R)}({\bf x},{\bf x}_{1};\epsilon+\omega)G^{0(R)}({\bf 
x}_{1},{\bf x};\epsilon)-
G^{0(A)}({\bf x},{\bf x}_{1};\epsilon)G^{0(A)}({\bf 
x}_{1},{\bf x};\epsilon-\omega)
\nonumber \\
&=&(-2e)T\{ \sum_{\epsilon_{n}\geq 0}\int \frac{d{\bf p}}{(2\pi)^{2}}
[G^{0(R)}({\bf p},\epsilon_{n})]^{2}
-\sum_{\epsilon_{n}\leq 0} \int \frac{d{\bf p}}{(2\pi)^{2}}
[G^{0(A)}({\bf p},\epsilon_{n})]^{2} \nonumber \\
&=&-e^{2}N(0)\varphi_{\omega}({\bf R})
\int d\xi \{ [{\cal P}\frac{1}{i\pi \xi}+\delta(\xi)] 
+[-{\cal P}\frac{1}{i\pi \xi}+\delta(\xi)]\}
\nonumber \\
&=&-2N(0)e^{2}\varphi_{\omega}({\bf R})\;,
\end{eqnarray}
and 
\begin{eqnarray}
\rho_{\omega}^{A}({\bf R})&=&-(\frac{e}{mi})\frac{1}{4\pi i}\int 
d\epsilon (\tanh \frac{\epsilon}{2T}-\tanh \frac{\epsilon-\omega}{2T}) 
\int d{\bf x}^{\prime\prime}  (-e\varphi_{\omega}({\bf 
x}^{\prime\prime})+ e{\bf A}_{\omega}({\bf x}^{\prime\prime})\cdot {\bf 
p}_{{\bf x}^{\prime\prime}})\nonumber \\
&&\times [G^{0(R)}({\bf R}+{\bf r}/2,{\bf 
x}^{\prime\prime};\epsilon)G^{0(A)}({\bf x}^{\prime\prime},{\bf R}-{\bf 
r}/2;\epsilon)]_{{\bf r}\rightarrow 0}
\nonumber \\
&=&(-2e)\frac{1}{4\pi i} \int d\epsilon 
(-\Gamma^{+}_{\omega}({\bf R},\epsilon))\int \frac{d{\bf p}}{(2\pi)^{2}}
G^{0(R)}(\epsilon,{\bf p})G^{0(A)}(\epsilon,{\bf p})
\nonumber \\
&=&-\frac{iN(0)e\tau_{1}}{2}\int d\epsilon [\Gamma_{\omega}^{+}({\bf 
R},\epsilon)+\Gamma_{\omega}^{-}({\bf 
R},\epsilon)]
\nonumber \\
&=&2e^{2}N(0)\tilde{\varphi}_{\omega}({\bf R}) \;.
\end{eqnarray}

After an inverse Fourier transform, we have the charge density in real 
time space
\begin{equation} 
\rho({\bf R},t)=2e^{2}N(0)[\tilde{\varphi}({\bf R},t)-
\varphi({\bf R},t)]\;.
\label{EQ:TDGL-Q}
\end{equation}
From Eqs.~(\ref{EQ:TDGL-J}), (\ref{EQ:TDGL-Q}), and (\ref{EQ:TDGL-DF}) 
follows the continuity equation $\nabla \cdot {\bf J}+\partial 
\rho/\partial t=0$. 

\section{Discussions and Summary}
Combined with the Maxwell equations, which couple ${\bf A}$ and $\varphi$
with ${\bf J}$ and $\rho$, Eqs.~(\ref{EQ:TDGL-OP}), 
(\ref{EQ:TDGL-J}), (\ref{EQ:TDGL-Q}), 
together with Eqs.~(\ref{EQ:TDGL-DF})  
constitute a complete set of coupled 
time-dependent Ginzburg-Landau equations, which are our main results. 
Several features of the above results deserve special 
attention: It is well known that depairing of $s$-wave 
superconductors are due only to magnetic impurities.
However, nonmagnetic impurities can have direct depairing 
effects on unconventional $d$-wave pairing state.
Similarly, the relaxation of the $s$-wave order parameter 
is influenced only by magnetic impurities.
Therefore, the magnetic impurities as pair-breakers are essential
in the derivation of the corresponding TDGL equations for conventional 
$s$-wave superconductors.~\cite{Gor68} 
However, nonmagnetic impurities acting as depairing centers 
can directly affect the relaxation of the $d$-wave order parameter.   
Interestingly, for a mixed $d$- and $s$-wave symmetry superconductor 
with a high concentration of magnetic and nonmagnetic impurities  
such that $\tau_{1}T_{c}\ll 1$ and $\tau_{s}T_{c}\ll 1$, we have 
$\gamma_{d}^{-1}\approx \tau_{1}$, $\gamma_{s}^{-1}\approx \tau_{s}$. 
In this limit, the TDGL equations for the order parameters,
Eqs.~(\ref{EQ:TDGL1}) and (\ref{EQ:TDGL2}), become 
\begin{equation}
-\tau_{s}[\frac{\partial}{\partial t}
+2ie\tilde{\varphi}({\bf R},t)] \Delta_{s}^{*}({\bf R},t) 
=\frac{\delta f({\bf R},t)}{\delta \Delta_{s}}\;,
\label{EQ:TDGL1a}
\end{equation} 
\begin{equation}
-\tau_{1}[\frac{\partial}{\partial t}
+2ie\tilde{\varphi}({\bf R},t)]
\Delta_{d}^{*}({\bf R},t) 
=\frac{\delta f({\bf R},t)}{\delta \Delta_{d}}\;,
\label{EQ:TDGL2a}
\end{equation}  
where the coefficients $\alpha_s$,$\alpha_d$, and $\chi_{m,n}$ can also be 
simplified, but are not explicitly given here.

These set of TDGL equations valid under the strong gaplessness 
conditions are similar in form to that postulated
phenomenologically~\cite{ADB97} except that the 
relaxation parameters obtained here are $\gamma_{s}$ ($=\tau_{s}^{-1}$) 
and $\gamma_{d}$ ($=\tau_{1}^{-1}$) and the usual scalar potential 
$\varphi$ is replaced by the electro-chemical potential $\tilde{\varphi}$.  
Therefore, the phenomenological TDGL equations are at most valid when
the superconductor is very dirty with also a high concentration of 
magnetic impurities.
If the superconductor is doped only with high density of nonmagnetic 
impurities ($\tau_{s}\gg \tau_{1}$), the TDGL 
equation~(\ref{EQ:TDGL2}) for $d$-wave component is reduced to 
Eq.~(\ref{EQ:TDGL2a}) while the relaxation parameter involved in the 
equation for  $s$-wave component becomes    
$\gamma_{s}^{-1} \approx \pi/4T_{c}$. 
In this case, $\gamma_{s}\ll 
\gamma_{d}$  and the TDGL equations for both components are 
quite asymmetric. Of particular interest, if $T_{cs}<T<T_{cd}$, due to a 
mixed gradient coupling of the $s$- and $d$-wave components, the $s$-wave 
order parameter with four-lobe structure is induced near the $d$-wave 
vortex core, and the overall structure of an individual vortex is 
fourfold symmetric. Numerical simulation,~\cite{ADB97} where the same 
relaxation rate ($\gamma_{s}=\gamma_{d}=\gamma$) was assumed 
for two components,  showed an intrinsic contribution to the Hall angle 
caused by the lack of complete rotational symmetry in $d$-wave 
superconductivity. In the case $\tau_{s}\gg \tau_{1}$, 
we could have $\gamma_{s}\ll \gamma_{d}$ and the $d$-wave order 
parameter relaxes much faster than the $s$-wave component.
Under this condition we expect that the $s$-wave component will not be 
able to follow the motion of the $d$-wave vortex and novel phenomenon 
may appear in the flux dynamics. Even when $\tau_1T_c$ and $\tau_sT_c$ 
are both small, the condition $\gamma_{d}=\gamma_{s}$ 
used in Ref.~\cite{ADB97} would require the assumption that the 
non-spin-flip interaction $U_1=0$, which as judged from the studies on 
conventional $s$-wave superconductors,~\cite{Gor68} may well be not 
justifiable. 
 
In summary, we have derived the TDGL equations for superconductors with 
mixed $d$-wave and $s$-wave symmetry assuming a weak gapless condition for
both types of order parameters. From this derivation, the unknown 
coefficients for the TDGL equations postulated phenomenologically have 
been ascertained.  This set of TDGL equations can be used as 
the starting point for the study of the vortex dynamics in superconductors 
with the mixed $d$- and $s$-wave symmetry, or even extended to study other
transport coefficients. 
In particular, the issue of how the dynamic properties of vortices 
are influenced by the admixture of an induced $s$-wave component with 
the dominant $d$-wave component of the order parameter as 
well as their different responses to the impurity scatterings can be 
studied systematically.  
The TDGL equations for $d$-wave superconductors with on-site $s$-wave 
repulsive interaction can be similarly obtained by using the 
Pad\'{e} approximation,~\cite{RXT95} and we find that the main conclusion 
of the present paper still remains unchanged. This result together 
with a detailed derivation will be presented elsewhere. Finally, we 
remark that the present derivation has not included the effects of 
electron-electron (actually hole-hole), electron-phonon, and 
electron-``any magnetic excitation'' scatterings, which might be more 
important in high-$T_c$ superconductors than in conventional low-$T_c$ 
superconductors.
Whereas such inelastic scatterings are far from being easy to incorporate
within the present framework, we think that their dominant qualitative
and perhaps semi-quantitative effects can be
taken into account phenomenologically by adding a term $1/\tau_E$
to the diffusion operator $\partial/\partial t - D\nabla^2$, where
$\tau_E$ stands for an inelastic relaxation time (assuming that the weak
gaplessness conditions are still satisfied).  
Consistent with such an approach one should regard $\tau_1$ and $\tau_s$ 
as effective, including also some effects of the inelastic scatterings. 

\acknowledgments
One of the authors (J.-X.Z.) thanks L. Sheng for helpful discussions.
This work was supported by Texas Center for Superconductivity 
at the University of Houston and by the Robert A. Welch Foundation.

\begin{figure}
\caption{Ladder-type diagram leading to $I(\omega,{\bf k})$. 
The momenta and frequencies for the solid (electron) lines in the upper part 
are ${\bf p}$ and $\epsilon$, and those for the solid lines in the lower 
part are ${\bf p}-{\bf k}$ and $\epsilon-\omega$. 
} 
\label{FIG:I}
\end{figure}

\begin{figure}
\caption{Impurity-averaged  diagrams leading to the diffusion 
equation for $\Gamma^{+}$. The thick wavy lines correspond to 
$I(\omega,{\bf k})$ shown in Fig.~\ref{FIG:I}. 
The thin wavy line corresponds to the vertex interaction with 
the electromagnetic field. The triangle represents the order parameter. 
}
\label{FIG:GAMMA}
\end{figure}

\begin{figure}
\caption{Impurity-averaged  diagrams for kernel $Q_{1}$. 
$\tilde{\Delta}$ and $\tilde{\Delta}^{*}$ are both the   
vertex-renormalized order parameters in the upper part. }
\label{FIG:Q1}
\end{figure}

\begin{figure}
\caption{Impurity-averaged  diagrams for kernel $Q_{3}$.
$\tilde{\Delta}$ is  the
vertex-renormalized order parameter in the upper part and 
$\tilde{\Delta}^{*}$ the vertex-renormalized order parameter in the lower 
part. }
\label{FIG:Q3}
\end{figure}

\begin{figure}
\caption{Impurity-averaged diagrams leading to the anomalous part in the 
TDGL equation for order parameter.
$\Gamma^{\pm}$ are given by the type of diagrams 
shown in Fig.~\ref{FIG:GAMMA}.
} 
\label{FIG:ANOM-OD}
\end{figure}

\begin{figure}
\caption{Impurity-averaged diagrams leading to the anomalous current 
density. 
} 
\label{FIG:ANOM-J}
\end{figure}

\end{document}